\documentclass[aps,prl,twocolumn,superscriptaddress]{revtex4-2}
\usepackage{amsmath}
\usepackage{mathtools}
\usepackage{amssymb}
\usepackage{mathrsfs}
\usepackage{dsfont}
\usepackage[pdftex,colorlinks=true,urlcolor=blue,linkcolor=blue,citecolor=blue,breaklinks=true,bookmarks=false]{hyperref}
\usepackage{graphicx}
\usepackage{float}
\usepackage{xcolor}

\newcommand{\push}{\hspace{0.05cm}}
\newcommand{\pull}{\hspace{-0.05cm}}
\newcommand{\spull}{\hspace{-0.025cm}}

    \usepackage{multibib} 

    \usepackage{array}   
    \newcolumntype{R}{>{$}r<{$}}
    \newcolumntype{L}{>{$}l<{$}}
    \usepackage{capt-of}  

    \usepackage{titlesec}
    \titleformat*{\section}{\large
    \bfseries}
    \titleformat*{\subsection}{\bfseries}
    \titleformat*{\subsubsection}{\itshape}

    \makeatletter
    \def\l@f@section{%
    \addpenalty{\@secpenalty}%
    \addvspace{0.9em plus\p@}
    }
    \def\l@subsection{%
    \l@@sections{section}{subsection}
    }
    \def\l@f@subsection{%
    \addpenalty{\@secpenalty}%
    \addvspace{0.2em plus\p@}
    }
    \def\l@subsubsection#1#2{} 
    \makeatother

\begin{document}

\title{Density-Matrix Renormalization Group for Continuous Quantum Systems}

\author{Shovan Dutta}
\affiliation{T.C.M. Group, Cavendish Laboratory, University of Cambridge, JJ Thomson Avenue, Cambridge CB3 0HE, United Kingdom\looseness=-1}
\affiliation{Laboratory of Atomic and Solid State Physics, Cornell University, Ithaca, New York 14853, USA}
\author{Anton Buyskikh}
\affiliation{Department of Physics and SUPA, University of Strathclyde,  Glasgow G4 0NG, United Kingdom}
\author{Andrew J. Daley}
\affiliation{Department of Physics and SUPA, University of Strathclyde,  Glasgow G4 0NG, United Kingdom}
\author{Erich J. Mueller}
\affiliation{Laboratory of Atomic and Solid State Physics, Cornell University, Ithaca, New York 14853, USA}

\date{\today}

\begin{abstract}
We introduce a versatile and practical framework for applying matrix product state techniques to continuous quantum systems. We divide space into multiple segments and generate continuous basis functions for the many-body state in each segment. By combining this mapping with existing numerical Density-Matrix Renormalization Group routines, we show how one can accurately obtain the ground-state wave function, spatial correlations, and spatial entanglement entropy directly in the continuum. For a prototypical mesoscopic system of strongly-interacting bosons we demonstrate faster convergence than standard grid-based discretization. We illustrate the power of our approach by studying a superfluid-insulator transition in an external potential. We outline how one can directly apply or generalize this technique to a wide variety of experimentally relevant problems across condensed matter physics and quantum field theory.
\end{abstract}

\maketitle

Few computational techniques have improved our understanding of strongly-correlated quantum systems as much as the Density-Matrix Renormalization Group (DMRG) \cite{schollwock2005density}. Introduced for calculating ground states of spin chains \cite{white1992density}, DMRG
takes advantage of the entanglement properties of many physical states \cite{eisert2010colloquium} to efficiently truncate the Hilbert space, approximating the many-body wave function as a variational Matrix Product State (MPS) \cite{schollwock2011density}. It has been successfully generalized for time evolution \cite{paeckel2019time} and is the method of choice for simulating {\it discrete} one-dimensional (1D) quantum systems \cite{baiardi2020density}, with promising extensions to higher dimensions \cite{verstraete2004renormalization} and other tensor networks \cite{orus2019tensor}. However, despite wide-ranging potential applications \cite{cazalilla2011one, guan2013fermi, bloch2008many, tang2018thermalization, eigen2018universal, erne2018universal, kunkel2021detecting}, attempts to generalize DMRG to continuous systems have encountered substantial difficulties. Here, we put forward a new framework that addresses this long-standing challenge.

The very formalism of DMRG, and other tensor network approaches, is predicated on having a lattice. To apply the technique to a continuous system, one must define a network of local Hilbert spaces.  The naive approach involves replacing the continuum with a lattice  \cite{sugihara2004density, weir2010studying, milsted2013matrix, stoudenmire2012one, stoudenmire2017sliced, knap2014quantum, bellotti2017comparing}. Unfortunately, this strategy scales poorly with the number of grid points and displays convergence issues in systems with multiple length scales \cite{bellotti2017comparing, dolfi2012multigrid}, requiring optimization on successively finer grids \cite{dolfi2012multigrid}, which becomes intractable for vanishingly small grid spacing \cite{schollwock2007progress}. Alternatively, by taking this continuum limit one can derive a field-theoretic description, called continuous MPS (cMPS) \cite{verstraete2010continuous}, which has had considerable success for translationally-invariant systems \cite{ganahl2017continuous, draxler2013particles, quijandria2015continuous, stojevic2015conformal, rincon2015lieb, haegeman2010applying, chung2015matrix, chung2017multiple, draxler2017continuous}, but has severe limitations in the presence of inhomogeneities: Interpolation-based algorithms \cite{ganahl2017inhom, tuybens2021variational} suffer from instabilities unless starting from a preconverged initial state, obtained from (multi)grid optimization \cite{ganahl2018continuous}. Furthermore, unlike DMRG, these cMPS calculations are intrinsically nonlinear \cite{haegeman2013calculus, haegeman2017quantum}, limited to low entanglement \cite{ganahl2017inhom}, and do not usually conserve particle number \cite{verstraete2010continuous, maruyama2010continuous}.

In contrast, we partition a continuous system into multiple {\it segments} and choose a flexible set of basis functions in each segment to describe the local physics. This recipe turns the Hamiltonian into a sum over segments, with nearest-neighbor terms imposing continuity at the boundaries. This form can be minimized using standard DMRG routines \cite{itensor}, used as a local basis for other tensor network algorithms, or even used for neural-network based approaches \cite{neural}. Like a Hubbard Hamiltonian, one can easily incorporate symmetries \cite{mcculloch2007density} such as particle number, and the method works equally well for homogeneous and inhomogeneous systems, regardless of the initial state. For many segments and few basis functions, it reduces to discretizing on a grid; however, we show that for interacting bosons in a box one can optimize the local basis to gain faster convergence with a small number of segments. We illustrate the broad applicability of this technique by exploring the Mott-superfluid transition in a sinusoidal potential.

For concreteness, we consider a paradigmatic Hamiltonian describing bosons with contact interactions \cite{lieb1963exact} trapped in a box of length $L$,
\begin{equation}
    \hat{H} = \int_0^L \pull {\rm d}x \push \bigg[
    \frac{1}{2}\frac{{\rm d}\hat{\psi}^{\dagger}}{{\rm d}x} \frac{{\rm d}\hat{\psi}}{{\rm d}x} 
    + \frac{g}{2} \hat{\psi}^{\dagger} \hat{\psi}^{\dagger} \hat{\psi} \hat{\psi} 
    + V(x) \hat{\psi}^{\dagger} \hat{\psi}
    \bigg] \push ,
    \label{eq:hamiltonian}
\end{equation}
where $\smash{\hat{\psi}(x)}$ is the boson field operator, $g$ is the interaction strength, $V(x)$ is an external potential, and we have set $\hbar=m=1$, where $m$ is the boson mass. This 1D model is realized for atoms with tight transverse confinement \cite{Kinoshita2004, kinoshita2006quantum}, with optical box traps \cite{ eigen2018universal, erne2018universal} or atom chips \cite{van2010box, tajik2019designing, rauer2018recurrences}. Its physics depends crucially on the ratio of interaction and kinetic energies, set by the dimensionless parameter $\gamma \coloneqq gL/N$, $N$ being the particle number. When $V(x)=0$, the model has an exact Bethe-Ansatz solution \cite{gaudin1971boundary, batchelor20051d, hao2006ground}, but calculating spatial correlations is challenging except for $\gamma \ll 1$ \cite{carr2000stationary} and $\gamma \to \infty$ \cite{forrester2003painleve}. Thus, one typically resorts to low-energy approximations \cite{cazalilla2004bosonizing}.

To use standard DMRG techniques, one can discretize Eq.~\eqref{eq:hamiltonian} on a grid of spacing $\epsilon$, mapping $\smash{\hat{\psi}}$ to lattice bosons, $\smash{\hat{\psi}(x)} \to \smash{\hat{b}_i}/\sqrt{\epsilon}$, and replacing derivatives by finite differences, which gives
\begin{equation}
    \hspace{-0.15cm}\hat{H} \approx -\frac{1}{2\epsilon^2} \sum_{\langle i,j \rangle} \hat{b}_i^{\dagger} \hat{b}_j 
    + \sum_{j=1}^{M-1} \frac{g}{2\epsilon} \push \hat{b}_j^{\dagger} \hat{b}_j^{\dagger} \hat{b}_j \hat{b}_j 
    + \Big(V_j + \frac{1}{\epsilon^2}\Big) \hat{b}_j^{\dagger} \hat{b}_j \push ,
    \label{eq:hamildisc}
\end{equation}
where $\langle i,j \rangle$ denotes nearest neighbors and $M-1$ is the number of grid points in the bulk, $L=M\epsilon$; see Ref.~\cite{ganahl2018continuous} for an alternative mapping to hard-core bosons. The continuum limit is obtained for $M \to\infty$.

Instead, we divide the box into $M$ finite segments with
\begin{equation}
    \hat{\psi}(x) = \sum\nolimits_{j=1}^M \Box_{X_{j-1},X_j} (x) \push \hat{\psi}(x) \push ,
    \label{eq:partition}
\end{equation}
where the box function $\Box_{a,b}(x)$ vanishes unless $a<x<b$; $\Box_{a,b}(x) \coloneqq \theta(x-a)-\theta(x-b)$, with $\theta$ denoting the unit step function. Thus, $X_j$ is the boundary between the $j$-th and the $(j+1)$-th segments, with $X_0 = 0$ and $X_M = L$. Substituting Eq.~\eqref{eq:partition} into Eq.~\eqref{eq:hamiltonian} and keeping track of delta functions, we find (see Supplement \cite{SuppMat})
\begin{equation}
    \hat{H} = \sum\nolimits_{j=1}^M \hat{K}_j + \hat{U}_j + \hat{P}_j + \Lambda \sum\nolimits_{j=0}^M \hat{\Upsilon}_{j,j+1} \push ,
    \label{eq:discretesum}
\end{equation}
where $\hat{K}_j$, $\hat{U}_j$, and $\hat{P}_j$ are, respectively, the kinetic, interaction, and potential energies in the $j$-th segment, given by integrals between $X_{j-1}$ and $X_j$, and
\begin{equation}
    \hat{\Upsilon}_{j,j+1} \coloneqq \big[\hat{\psi}(X_j^{-}) - \hat{\psi}(X_j^+)\big]^{\dagger} \big[\hat{\psi}(X_j^{-}) - \hat{\psi}(X_j^+)\big]
    \label{eq:discontinuity}
\end{equation}
is a positive-semidefinite measure of the discontinuity between $x\to X_j^-$ and $x\to X_j^+$. We use hard-wall boundary conditions at the edge of our system, and in Eq.~\eqref{eq:discontinuity} define $\hat{\psi}(0^-) \coloneqq \hat{\psi}(L^+) \coloneqq 0$. The prefactor $\Lambda$ is a formally infinite energy penalty that projects onto continuous states. In our numerical calculations, we take $\Lambda$ to be finite, increasing it in consecutive DMRG cycles. This approach accelerates convergence because the system takes larger steps through phase space when $\Lambda$ is smaller.

\begin{figure}
	\includegraphics[width=1\columnwidth]{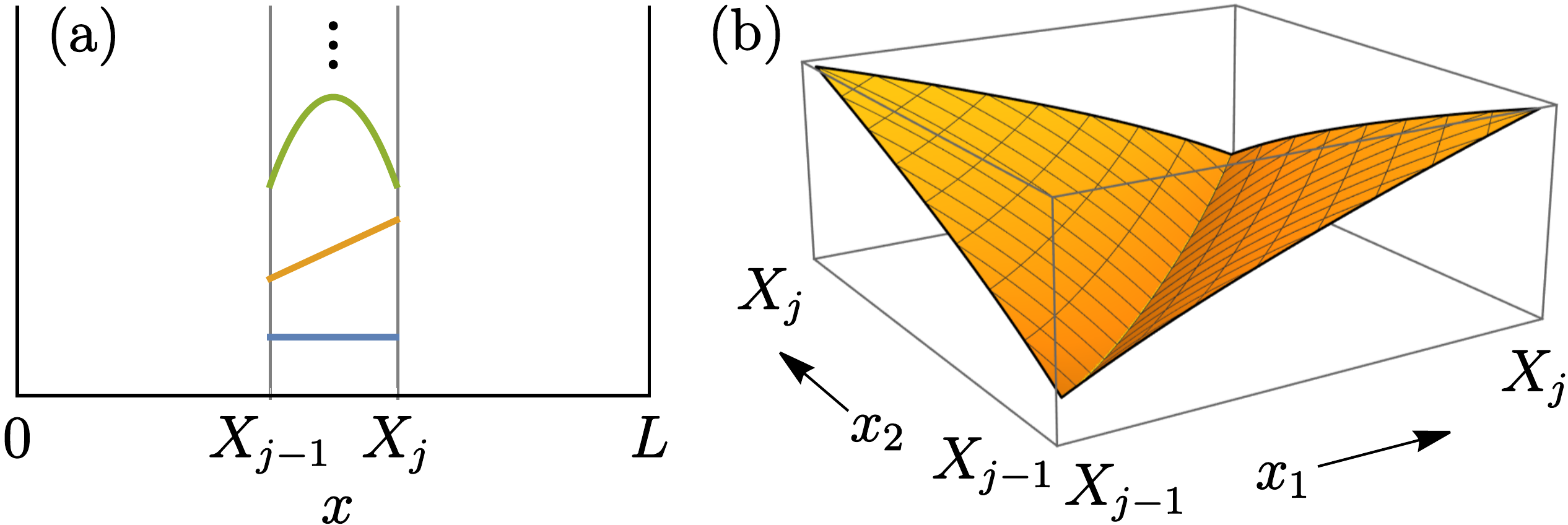}
	\centering
	\caption{Examples of (a) one-body and (b) two-body basis functions in a segment between $X_{j-1}$ and $X_j$. The cusp at $x_1 = x_2$ in (b) encodes the physics of contact interactions.}
	\label{fig:basisfns}
\end{figure}

Equation~\eqref{eq:discretesum} has the same form as a Hubbard model, with ``on-site'' and nearest-neighbor terms that can be expressed as a compact Matrix Product Operator (MPO) \cite{schollwock2011density}, acting as the input to a DMRG calculation. We select $n$-body basis functions $\smash{\phi^{(j)}_{n,k}(\Vec{r})}$ in each segment $j$, where $n=0,1,\dots$ up to some cutoff $n_{\text{max}} \leq N$, and $k$ labels the different states for a given $n$ (again with some cutoff). The construction of these basis functions is described below and examples of one- and two-particle states are shown in Fig.~\ref{fig:basisfns}. In contrast, for a lattice model as in Eq.~\eqref{eq:hamildisc}, the local bases are simply labeled by the number of particles on each site, $|0\rangle_j,\dots,|n_{\text{max}}\rangle_j$. Once our continuous bases are chosen, one finds the matrix elements as
\begin{flalign*}
    \spull \big\langle \phi^{(j)}_{n \spull-\spull 1,k} \big| \hat{\psi}(x) \big| \phi^{(j)}_{n,k^{\prime}} \big\rangle
    \spull &= 
    \spull\sqrt{n} \pull \int_{X_{j \spull-\spull 1}}^{X_j} \pull\pull\spull {\rm d}^{n \spull-\spull 1} r \push \phi^{(j)*}_{n \spull-\spull 1,k}(\vec{r}) \push \phi^{(j)}_{n,k^{\prime}} (x,\vec{r}) \push , &
\end{flalign*}
where $x \in [X_{j-1},X_j]$ and $n \geq 1$. Similar expressions for the matrix elements of $\smash{\hat{K}_j}$, $\smash{\hat{U}_j}$, and $\smash{\hat{P}_j}$ are derived in the Supplement \cite{SuppMat}. Note these operators conserve particle number and are thus block diagonal. If we choose the segments to have equal width, then the basis functions on different segments can be translations of one another, and the local matrices become independent of $j$.

We take the basis functions to be piecewise polynomials, i.e., for $X_{j-1}\leq x_1 \leq x_2 \leq \dots \leq x_n\leq X_j$,  $\smash{\phi^{(j)}_{n,k}(\vec{R}_{j-1}+\vec{r})} = \sum_{\mathbf{p}} \smash{A^{(j)}_{n,k,\mathbf{p}}} x_1^{p_1} \dots x_n^{p_n}$, and the other sectors are determined by symmetry under particle exchange. Here, $\mathbf{p}=\{p_1,\dots,p_n\}$ is a vector of the powers that appear in each monomial. As a practical strategy, we limit the maximum degree of the monomials: $p_1 + \dots + p_n \leq d_{\text{max}}$, and choose the coefficients $\smash{A^{(j)}_{n,k,\mathbf{p}}}$ so that the basis is orthonormal, $\langle \phi^{(j)}_{n,k} | \phi^{(j)}_{n,k^{\prime}} \rangle = \delta_{k,k^{\prime}}$. Given these constraints, we wish to construct polynomials that capture the low-energy physics with a minimum number of states. For example, the contact interaction in Eq.~\eqref{eq:hamiltonian} gives rise to a kink in the wavefunction wherever two particles coincide \cite{lieb1963exact}, $\partial_{x_i} \Psi ({x_i \to x_{i^{\prime}}^{+}}) - \partial_{x_i} \Psi ({x_i \to x_{i^{\prime}}^{-}}) = g \Psi ({x_i = x_{i^{\prime}}})$, and the numerics are more efficient if we include the same kink in the basis functions $\smash{\phi^{(j)}_{n,k}(\vec{r})}$. In the Supplement \cite{SuppMat}, we show how to construct generalizations of Legendre polynomials that possess these cusps. Calculating the local basis, and the matrix elements of the local operators, only needs to be done once and makes a negligible contribution to the computation time, which is dominated by the DMRG sweeps.

With this construction, the matrix elements of local operators become piecewise polynomial functions, of the form
$\spull \big\langle \phi^{(j)}_{n ,k} \big| \hat{\cal F}(x) \big| \phi^{(j)}_{m,k^{\prime}} \big\rangle \spull= \smash{\sum_{p=0}^{p_{\text{max}}}} \smash{F^{(j,p)}_{nk,mk^\prime}} (x - X_{j-1})^p $. Consequently, spatial correlations $\langle \hat{\mathcal{F}}^{\dagger}(x) \hat{\mathcal{F}}(x^{\prime}) \rangle$ can be expressed as piecewise polynomials, which one can evaluate at any point, once the matrices $\mathcal{F}^{j,p}$ are stored.
   
\begin{figure}
	\includegraphics[width=1\linewidth]{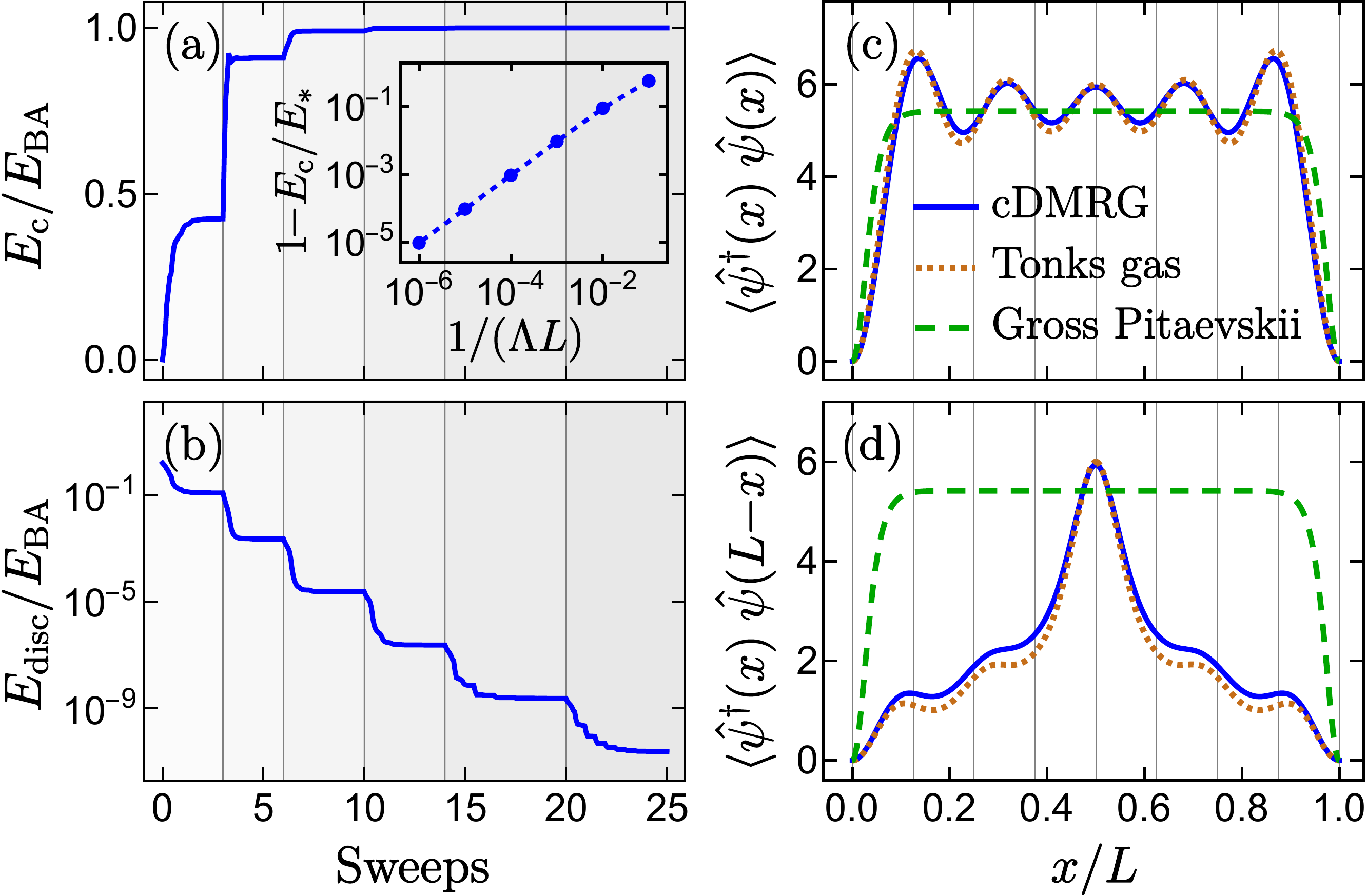}
	\centering
	\caption{Ground-state calculation for $N=5$ bosons with strong interactions ($\gamma = 50$) in a uniform box, divided into 8 segments. (a,b) Starting from a localized state, as the penalty $\Lambda$ is increased in powers of 10, the discontinuity falls as $1/\Lambda^2$ and the energy approaches the asymptotic value $E_*$, which is within $5\times 10^{-6}$ of the exact Bethe-Ansatz result $E_{\text{BA}}$. (The remaining discrepency is due to the finite basis set used.) Here, $E_c = \smash{\sum_j} \langle \hat{K}_j + \hat{U}_j \rangle$ and $E_{\text{disc}} \coloneqq (N/L) \smash{\sum_j} \langle \smash{\hat{\Upsilon}_{j,j+1}} \rangle$ [see Eqs.~\eqref{eq:discretesum}-\eqref{eq:discontinuity}]. The inset shows $E_c$ approaching $E_*$ as $1/\Lambda$. (c,d) Density and correlations in the ground state from continuous DMRG (cDMRG), showing Friedel oscillations similar to exact results for the Tonks gas ($\gamma \to \infty$) and far from a mean-field Gross-Pitaevskii theory. See supplement \cite{SuppMat} for a full description of the basis states and the DMRG parameters.
	}
	\label{fig:samplecalc}
\end{figure}

Figure~\ref{fig:samplecalc} shows a benchmark calculation for 5 strongly-interacting bosons in a uniform trap [$V(x)=0$], divided into $M=8$ equal segments with basis functions that can describe quartic variations, i.e., $d_{\text{max}}=4$. We initialize the particles in a discontinuous product state, where each segment contains either zero or one particle, and the single-particle wave function is uniform, hence $E_c \pull \coloneqq \smash{\sum_j} \langle \smash{\hat{K}_j} + \smash{\hat{U}_j} \rangle = 0$. We use the standard DMRG algorithm to minimize $\hat{H}$ in Eq.~\eqref{eq:discretesum} with a small penalty $\Lambda$. As shown in Figs.~\ref{fig:samplecalc}(a-b), $E_c$ increases with each sweep, and the discontinuity drops. After convergence, we sequentially increase $\Lambda$ by factors of 10, stopping when the discontinuity falls below a small threshold. For large $\Lambda$, the energy saturates at $E_c \approx E_* - \eta/\Lambda$ with constant $\eta$, from which one can robustly extrapolate $E_*$. Already with $M=8$, $E_*$ matches the ground-state energy from Bethe Ansatz \cite{batchelor20051d} to $5 \times 10^{-6}$. The density in Fig.~\ref{fig:samplecalc}(c) shows oscillations that are similar to those found in the Tonks gas, which would model the system for $\gamma\to\infty$ \cite{forrester2003painleve}. In that limit, these corrugations can be interpreted as the Friedel oscillations of a free Fermi gas \cite{hao2006ground}, which are not reproduced in mean-field theory \cite{carr2000stationary}. The single-particle correlator in Fig.~\ref{fig:samplecalc}(d) has a peak at small distances, and distinctive steps. The expected Luttinger-liquid power-law tail \cite{giamarchi2003quantum} is cut off by finite-size effects. Again, the result is similar to what one expects for a Tonks gas and is very different from the mean-field prediction.

\begin{figure}
	\includegraphics[width=1\columnwidth]{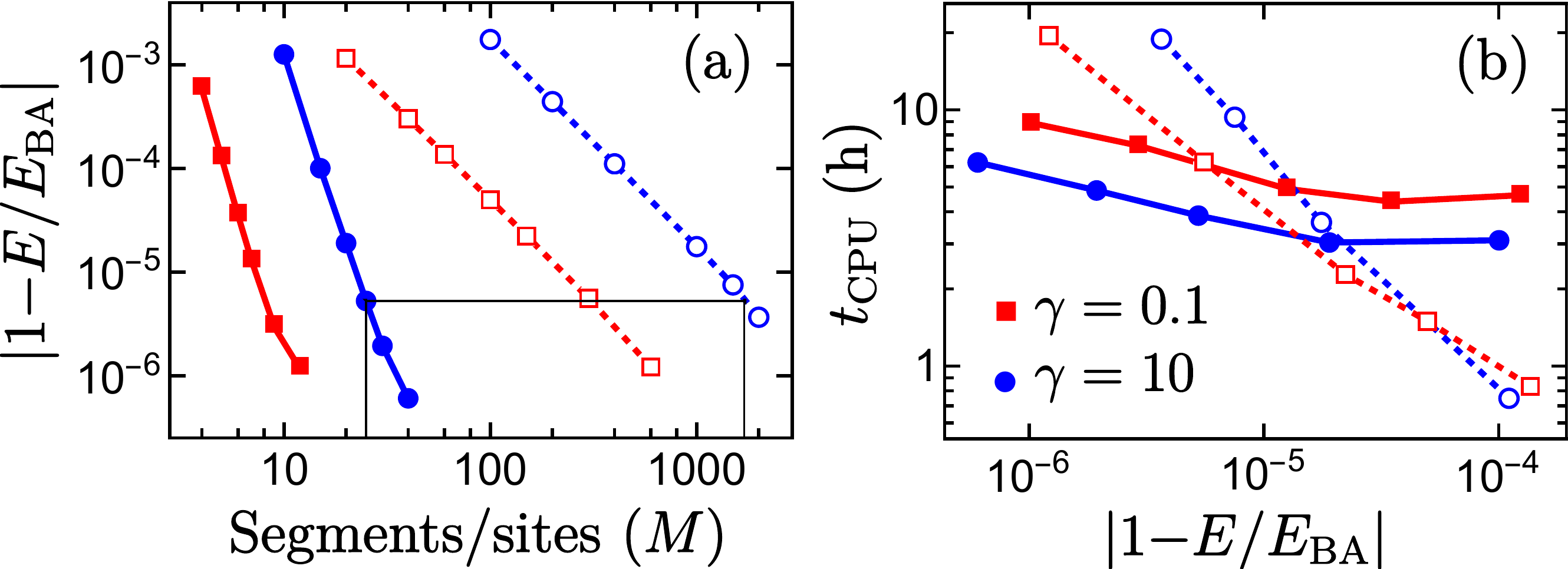}
	\centering
	\caption{(a) Relative error vs number of segments (grid points) and (b) CPU time vs relative error in finding ground states for $N=10$ using cDMRG (solid lines) and discretization on a grid (dotted lines). For cDMRG, $E$ is the extrapolated energy $E_* \pull > E_{\text{BA}}$ (see Fig.~\ref{fig:samplecalc}). We retained up to cubic basis functions in each segment, causing the error to fall off as $M^{-6}$, instead of $M^{-2}$ for discretization. Empirically, $t_{\text{CPU}} \sim |1-E/E_{\text{BA}}|^{-0.2}$ for cDMRG at small errors, whereas for discretization this exponent is $-0.75$ for $\gamma=0.1$ and $-1$ for $\gamma=10$. Note, $t_{\text{CPU}}$ was measured in seconds from the number of clock cycles during all DMRG sweeps on single quad-core CPUs. The saturation at large errors in (b) is due to larger bond dimensions \cite{SuppMat}.
	}
	\label{fig:benchmarking}
\end{figure}

Figure~\ref{fig:benchmarking} explores the performance of our algorithm, cDMRG, and compares it  with the grid-based discretization in Eq.~\eqref{eq:hamildisc}. We consider $N=10$ particles and piecewise cubic basis states ($d_{\text{max}}=3$). As illustrated by panel (a), as one refines the grid, the error in ground-state energy falls off as $M^{-2 d_{\text{max}}}$: The kinetic energy per segment $\langle \smash{\hat{K}_j} \rangle$ can be approximated up to that order. The standard discretization instead shows an error scaling as $M^{-2}$. Increasing $d_{\text{max}}$ allows one to achieve the same accuracy with fewer segments, at the cost of a larger local basis. The relationship between CPU time and accuracy is shown in Fig.~\ref{fig:benchmarking}(b). The traditional discretization is more efficient for low-accuracy calculations, where the smaller local Hilbert space is beneficial. Our algorithm uses fewer computational resources for high-accuracy calculations, where precise modeling of the local physics is crucial. The crossover point depends on the interaction strength: cDMRG is more efficient for strong repulsive interactions, which suppress the occupation of basis states containing larger numbers of particles.

Since the ground-state entanglement entropy of this system grows as $\ln N$ \cite{calabrese2011entanglement, herdman2016spatial, simon2002natural}, we find a linear rise in the DMRG bond dimension \cite{pollmann2009theory} with particle number, and the computation time roughly scales as $N^3$ \cite{schollwock2011density}. Our calculations were done using the ITensor library \cite{itensor}, using a two-site DMRG algorithm with a singular-value cutoff of $10^{-14}$, resulting in bond dimensions of order 100 (see Supplement \cite{SuppMat} for details).

\begin{figure}[!htb]
	\includegraphics[width=1\columnwidth]{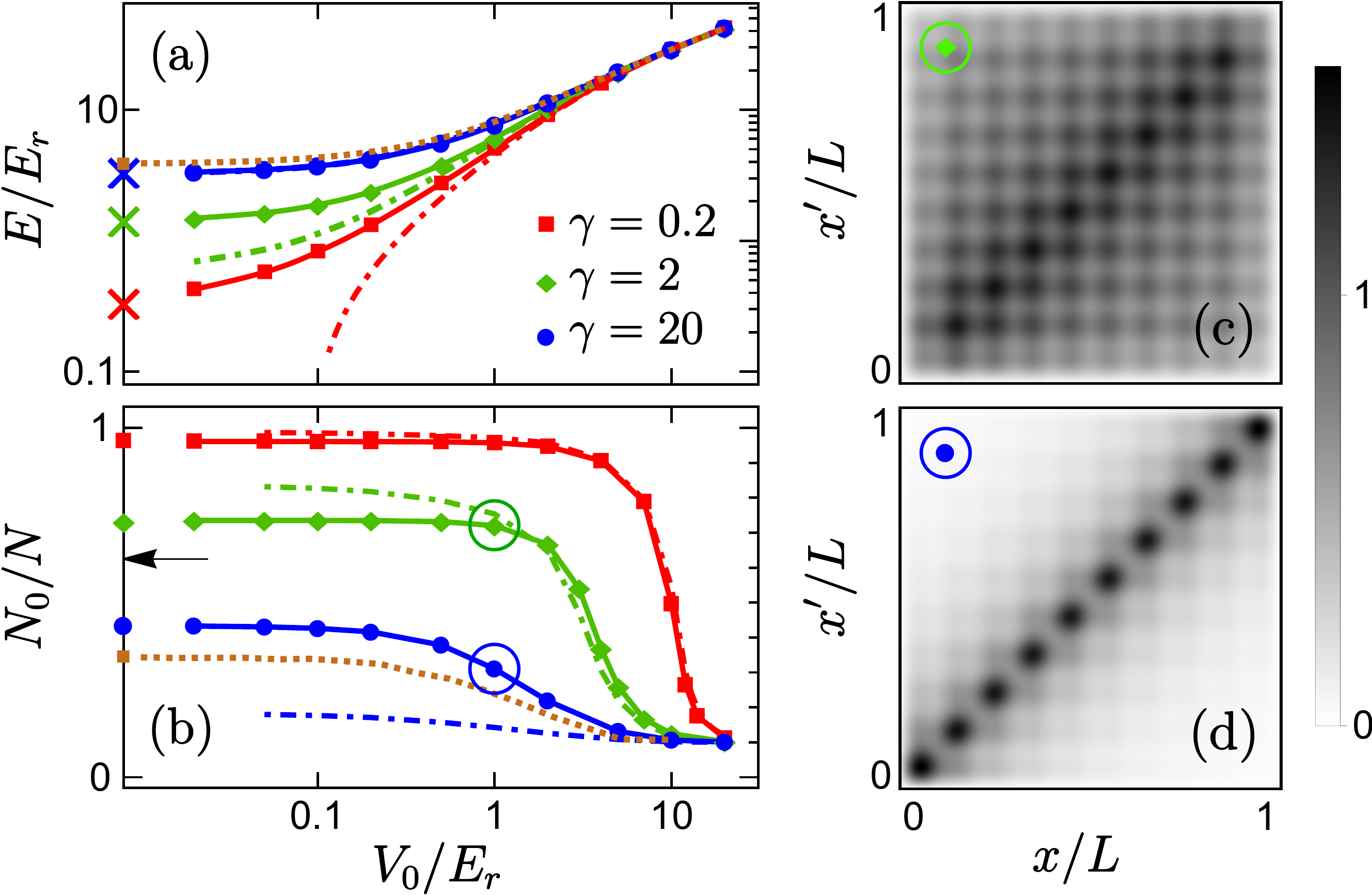}
	\centering
	\caption{(a) Ground-state energy and (b) condensate fraction for 10 bosons in 10 potential wells of depth $V_0$ using cDMRG with 20 segments and quartic basis functions (solid lines) and a tight-binding approximation (dash-dotted lines). Here, $N_0$ is the occupation of the single-particle ground state. Dotted lines in (a,b) and crosses in (a) show exact solutions for the Tonks gas and from Bethe Ansatz, respectively; the condensate fractions are found using Monte-Carlo integration \cite{SuppMat}. The arrow in (b) marks the $N_0$ that we find for $\gamma \approx 4.22$, when the ground state for $V_0 \to 0$ becomes Mott insulating. In our finite-size system this transition is a crossover. (c,d) Single-particle correlations, $\langle \smash{\hat{\psi}^{\dagger}(x)} \smash{\hat{\psi}(x^{\prime})}\rangle L/N$, in a superfluid and a Mott state, corresponding to the circled points in (b).
	}
	\label{fig:inhomogeneous}
\end{figure}

Next, we consider $V(x) = V_0 \cos^2(N_w \pi x/L)$, with $N_w$ potential wells between $0$ and $L$, which makes the system nonintegrable. There are two simple limits: (i) For $V_0 \gg E_r = N_w^2 \pi^2/(2L^2)$, where $E_r$ is the recoil energy, one can make a tight-binding approximation to reduce the problem to a Bose-Hubbard model with $N_w$ sites and slightly nonuniform parameters \cite{SuppMat}. (ii) For $\gamma \to \infty$, the system maps onto free fermions \cite{girardeau1960relationship}. Figure~\ref{fig:inhomogeneous}(a) shows how cDMRG reproduces these limits and smoothly connects the tight-binding and Bethe-Ansatz regimes  for all $\gamma$. For $N_w = N$, we find signatures of the superfluid-to-Mott-insulator transition \cite{stoferle2004transition} for both deep ($V_0 \gg E_r$) and shallow lattices ($V_0 \sim E_r$): As $\gamma$ is increased, the ground-state coherences localize, i.e., the algebraic variation of the correlation functions [Fig.~\ref{fig:inhomogeneous}(c)] becomes exponential [Fig.~\ref{fig:inhomogeneous}(d)], accompanied by a drop in the condensate fraction [Fig.~\ref{fig:inhomogeneous}(b)]. Similar to unbounded systems \cite{buchler2003commensurate, haller2010pinning} and those with periodic boundary conditions \cite{boeris2016mott, astrakharchik2016one}, the low-energy physics for $V_0 \to 0$ is described by a quantum sine-Gordon Hamiltonian \cite{giamarchi2003quantum}, which gives a Mott phase for $\gamma > \gamma_c \approx 3.5+7.5/N$ (see Supplement \cite{SuppMat}). Hence, the superfluid phase is found only for $\gamma < \gamma_c$ and sufficiently small $V_0$.

\begin{figure}
	\includegraphics[width=1\columnwidth]{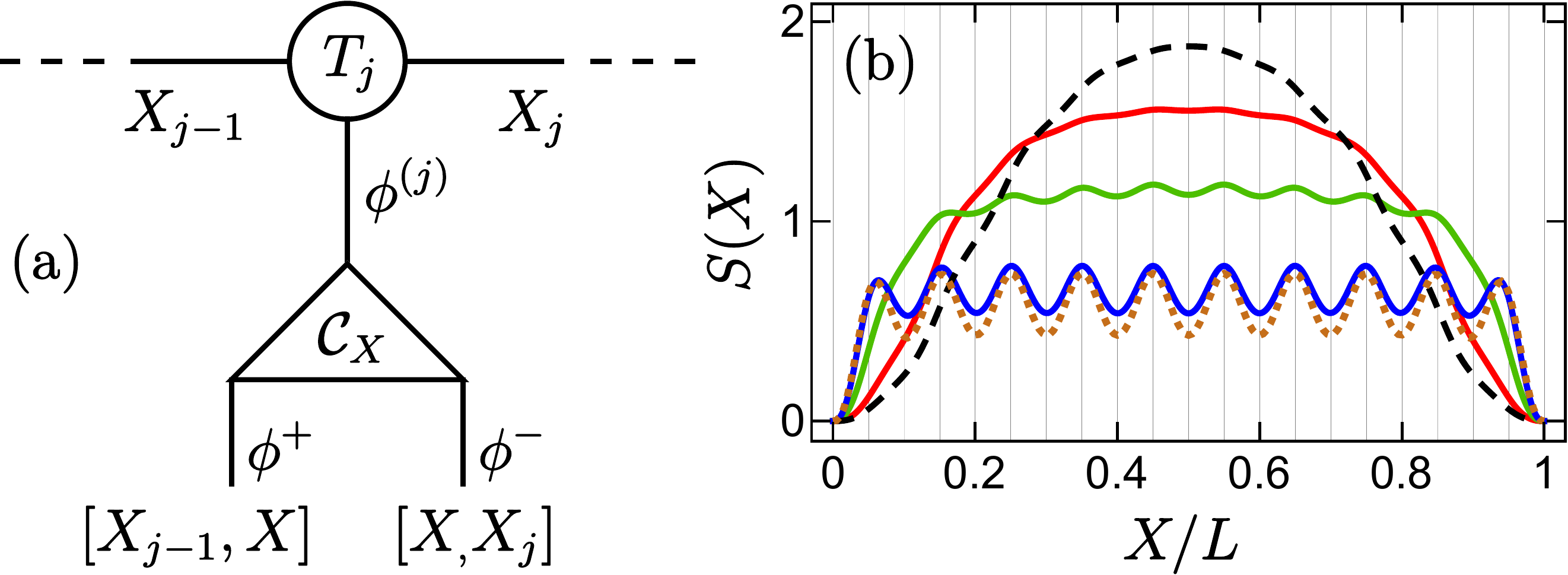}
	\centering
	\caption{(a) Schematic of how one can decompose a tensor $T_j$ of the MPS by a basis splitting $\smash{\phi^{(j)}} \pull \to \phi^+ \otimes \phi^-$ to calculate the entanglement between arbitrary bipartitions $[0,X]$ and $[X,L]$. (b) Ground-state entanglement entropy for 10 bosons in a shallow lattice ($V_0 = E_r$) with interactions $\gamma = 0.2$ (red), $2$ (green), $20$ (blue), using cDMRG with the same numerical parameters as in Fig.~\ref{fig:inhomogeneous}. We have split each segment at 20 intermediate points \cite{SuppMat}. Dashed and dotted curves show exact results for $\gamma=0$ \cite{simon2002natural} and $\gamma\to\infty$ \cite{calabrese2011entanglement}, respectively.
	}
	\label{fig:entanglement}
\end{figure}

A key feature of our approach is that one can compute the spatial bipartite entanglement entropy $S$ directly in the continuum, for which current understanding is limited \cite{kunkel2021detecting, calabrese2011entanglement, herdman2016spatial, simon2002natural, calabrese2004entanglement}. For bipartition at an arbitrary position $X \in (X_{j-1},X_j)$, we divide the $j$-th segment into left and right zones, with their own basis functions $\phi^{\pm}_{n,k}$, and write the original basis as a tensor product, $\smash{\phi^{(j)}_{n,k}} =$ $\smash{\sum_{n^{\prime},k^{\pm}} \mathcal{C}^{n,k}_{n^{\prime},k^+, k^-} \phi^+_{n^{\prime},k^+} \phi^-_{n-n^{\prime},k^-}}$ \cite{SuppMat}. Thus, one can express the local tensor $T_j$ of the MPS in the product basis [Fig.~\ref{fig:entanglement}(a)], and calculate $S(X)$ via a Schmidt decomposition \cite{schollwock2011density}.  If one only needs the entanglement at a segment boundary, the subdivision step can be skipped. Figure~\ref{fig:entanglement}(b) shows the entropy variation for bosons in a shallow lattice:  $V_0=E_r$.  At weak coupling, where there are large number fluctuations, the entropy is peaked about the center, as is characteristic of the critical superfluid phase \cite{calabrese2004entanglement}. In contrast, at strong coupling the entropy is largely flat, corresponding to the  short-range ``area law" entanglement expected in the Mott phase \cite{eisert2010colloquium}. Additionally, there are small wiggles that are related to Friedel oscillations [Fig.~\ref{fig:samplecalc}(c)]. This spatial variation can be measured in current experiments \cite{kunkel2021detecting} and used as a tool to characterize continuous phases \cite{calabrese2004entanglement}.

Finally, we give examples of how our framework can be applied more generally. First, for multicomponent bosons with contact interactions \cite{cazalilla2011one, rauer2018recurrences}, one can partition each component $\sigma$ into the same segments, altering Eq.~\eqref{eq:discretesum} to
\begin{equation}
    \hat{H} = \sum\nolimits_{j} \Big[\sum\nolimits_{\sigma} \hat{K}^{\sigma}_j + \hat{P}^{\sigma}_j + \Lambda \hat{\Upsilon}^{\sigma}_{j,j+1} + \sum\nolimits_{\sigma,\sigma^{\prime}} \hat{U}^{\sigma,\sigma^{\prime}}_j \Big] ,
    \label{eq:mixture}
\end{equation}
which still has only on-site and nearest-neighbor terms and thus the same MPO bond dimension. However, the basis functions will carry additional labels to accommodate all the components, e.g., $\smash{\phi^{(j)}_{n^{\uparrow},n^{\downarrow},k}}$ for $\sigma=\uparrow,\downarrow$. The pairwise interaction strengths will determine the corresponding kinks in the basis functions.

Second, multicomponent fermions \cite{guan2013fermi} can be treated on an {\it equal} footing as bosons, only using different basis functions for the segments. In particular, one replaces the cusp condition with the requirement that a basis function must vanish whenever two fermions in the same spin state coincide, $x^{\sigma}_i = x^{\sigma}_{i^{\prime}}$. This is the only distinction from bosons. One can clearly also work with mixtures of fermions and bosons \cite{guan2013fermi}.

Third, long-range interactions will couple pairs of segments, changing $\sum_j \hat{U}_j$ to $\sum_{j,j^{\prime}} \hat{U}_{j,j^{\prime}}$ in Eq.~\eqref{eq:discretesum}. The simplest case is that of an exponential decay \cite{rincon2015lieb}, $u(x,x^{\prime}) = e^{-\kappa |x - x^{\prime}|}$, which yields $\hat{U}_{j,j^{\prime}} = e^{\kappa \delta (j-j^{\prime})} \smash{\hat{V}^{+}_j \hat{V}^-_{j^{\prime}}}$ for segments of equal width $\delta$ and $j<j^{\prime}$, where $\hat{V}^{\pm}_j$ are weighted averages of the density in segment $j$. Such exponential terms increase the MPO bond dimension of $\hat{H}$ only by 1 \cite{schollwock2011density}. However, one often wants to model power-law interactions, $u(x,x^{\prime}) = f(|x-x^{\prime}|) |x-x^{\prime}|^{-\nu}$ where $f(|x|) \to 1$ for large $|x|$. While this form itself does not lead to a compact MPO, one can accurately approximate the power law by a sum of relatively few exponentials \cite{pirvu2010matrix, crosswhite2008applying} and further compress the resulting MPO \cite{frowis2010tensor}. Optimal basis functions can be estimated from the two-body problem as well as exact solutions for $\nu=2$ \cite{sutherland1971quantum}, which can serve as a benchmark for the nonintegrable cases such as dipolar interactions \cite{baranov2012condensed, browaeys2020many} and Coulomb repulsion \cite{schmidt2018optical}.

In summary, we have demonstrated a much needed versatile approach that opens up practical applications of DMRG to many-body continuous systems. This cDMRG technique uses spatial partitioning to map the problem to discrete DMRG, seemlessly integrating with existing routines.  Nonetheless, the wave function, correlations, and entanglement are obtained directly in the continuum. We have shown cDMRG gives accurate and efficient results for interacting bosons with or without an external potential, and outlined how it would apply to other setups of current interest with little modification. By using physically motivated basis functions, we are able to obtain fast convergence with a limited number of segments, avoiding the need for multiscale optimization \cite{dolfi2012multigrid} for mesoscopic systems. Nonetheless, cDMRG can work in conjunction with such approaches, if necessary, for very dilute gases or for finite-size scaling; the main ingredient for multiscale approaches is a protocol to map the wave function from a coarser to a finer basis, which we already implemented to calculate the entanglement entropy in Fig.~\ref{fig:entanglement} (see Supplement for details \cite{SuppMat}). Although we have focused on ground states, our approach can be combined with existing techniques for time evolution, such as the time-dependent variational principle \cite{haegeman2016unifying}. Since the latter conserves energy at each time step, a large energy penalty $\Lambda$ in Eq.~\eqref{eq:discretesum} would ensure that one evolves in the manifold of continuous states. This will enable exciting applications to unsolved nonequilibrium problems such as prethermalization of strongly-interacting bosons \cite{tang2018thermalization, eigen2018universal}, domain wall instability in Fermi superfluids \cite{lu2012expansion, bolech2012long, dutta2017collective}, and false vacuum decay in cosmology \cite{ng2021fate}. Thus, we believe cDMRG will find wide usage across condensed matter, quantum field theory, and quantum chemistry \cite{baiardi2020density}.

An open-source code is available at \href{https://github.com/Shovan-Physics/cDMRG}{https://github.com/Shovan-Physics/cDMRG}.

We thank Fran\c{c}ois Damanet and Stuart Flannigan for useful discussions. This work was supported by the Engineering and Physical Sciences Research
Council Programme
Grant DesOEQ (EP/P009565/1), and the National Science Foundation Grants PHY-1806357 and
PHY-2110250.


\begingroup
\renewcommand{\addcontentsline}[3]{}
\renewcommand{\section}[2]{}

\endgroup

\onecolumngrid
\clearpage
\renewcommand{\baselinestretch}{1.2}\normalsize
\begin{center}
\textbf{\large Supplemental Material:\\
Density-Matrix Renormalization Group for Continuous Quantum Systems}
\end{center}

    \renewcommand{\thefigure}{S\arabic{figure}}
    \renewcommand{\theHfigure}{S\arabic{figure}}
    \renewcommand{\theequation}{S\arabic{equation}}
    \renewcommand{\thetable}{S\arabic{table}}
    \renewcommand{\thesection}{S\Roman{section}}
    \renewcommand{\bibnumfmt}[1]{[S#1]}
    \renewcommand{\citenumfont}[1]{S#1}

    \setcounter{equation}{0}
    \setcounter{figure}{0}
    \setcounter{table}{0}
    \setcounter{page}{1}
    \setcounter{section}{0}
    \setcounter{secnumdepth}{3}
    \makeatletter

	\renewcommand{\baselinestretch}{1}\normalsize
    \tableofcontents
    \renewcommand{\baselinestretch}{1.25}\normalsize
    \normalfont

\section{Derivation of the partitioned Hamiltonian}
\label{suppsec:Hpart}

We consider the Hamiltonian in Eq.~\eqref{eq:hamiltonian} of the main text, $\hat{H} = \int_0^L {\rm d}x \left[\hat{\mathcal{K}}(x) + \hat{\mathcal{U}}(x) + \hat{\mathcal{P}}(x)\right]$, where 
\begin{align}
    \hat{\mathcal{K}}(x) &= 
    \frac{1}{2}\;\left( \frac{{\rm d}}{{\rm d}x}\hat{\psi}^{ \dagger }(x)\right) \push
    \left(\frac{{\rm d}}{{\rm d}x}\hat{\psi}(x)\right) \;,
    \label{suppeq:K}\\
    \hat{\mathcal{U}}(x) &= 
    \frac{g}{2}\; \hat{\psi}^{\dagger}(x) \push \hat{\psi}^{\dagger}(x) \push \hat{\psi}(x) \push \hat{\psi}(x) \;,
    \label{suppeq:U}\\
    \hat{\mathcal{P}}(x) &= 
    V(x)\; \hat{\psi}^{\dagger}(x) \push \hat{\psi}(x) \;
\end{align}
denote the kinetic, interaction, and potential energy densities, respectively. We divide the $x$ axis into $M$ continuous segments with boundaries at $\{X_j\}$. Since the wave function is everywhere finite, the interaction and potential energies reduce to sums over the individual segments, $\hat{U} = \sum_j \hat{U}_j$ and $\hat{P} = \sum_j \hat{P}_j$, where
\begin{align}
    \hat{U}_j = \int_{X_{j-1}}^{X_j} \pull\pull {\rm d}x \; \hat{\mathcal{U}}(x)\;, \quad \text{and} \quad
    \hat{P}_j = \int_{X_{j-1}}^{X_j} \pull\pull {\rm d}x \; \hat{\mathcal{P}}(x) \;.
    \label{suppeq:UjandPj}
\end{align}
However, the kinetic energy $\hat{K}$ will diverge if $\hat{\psi}(x)$ has a discontinuity across any of the boundaries, which gives rise to additional terms in the Hamiltonian. To see this, we write $\hat{\psi}(x)$ close to the boundary at $X_j$ as
\begin{equation}
    \hat{\psi}(x) = 
    \big[ 1 - \theta(x - X_j) \big] \push \hat{\psi}_{<}(x) +
    \theta(x - X_j) \push \hat{\psi}_{>}(x)
\end{equation}
where $\theta$ is the unit step function and $\hat{\psi}_{< (>)}(x)$ denotes the part of $\hat{\psi}$ to the left (right) of $X_j$. Thus, one obtains
\begin{equation}\frac{{\rm d}}{{\rm d}x}
    \hat{\psi}(x) = 
    \delta(x - X_j) \push \big[ \hat{\psi}_{>}(x) - \hat{\psi}_{<}(x) \big] +
    \big[ 1 - \theta(x - X_j) \big] \push \left(\frac{{\rm d}}{{\rm d}x}\hat{\psi}_{<}(x)\right) +
    \theta(x - X_j) \left(\push\frac{{\rm d}}{{\rm d}x} \hat{\psi}_{>}(x)\right) \;,
    \label{suppeq:psiprime}
\end{equation}
where $\delta$ is the Dirac delta function. Substituting this expression into Eq.~\eqref{suppeq:K} and using the sifting property of the delta function yield the following contribution to $\hat{K}$ from the boundary,
\begin{align}
    \nonumber
    \int_{X_j^-}^{X_j^+} \pull\pull {\rm d}x \; \hat{\mathcal{K}} = &\; 
    \Lambda\push 
    \big[ \hat{\psi}(X_j^+) - \hat{\psi}(X_j^-) \big]^{\dagger} 
    \big[ \hat{\psi}(X_j^+) - \hat{\psi}(X_j^-) \big]
    \\[-0.2cm]
    & + \frac{1}{2} \push
    \big[ \hat{\psi}(X_j^+) - \hat{\psi}(X_j^-) \big]^{\dagger} 
    \left[ (1 - \theta_0) \push\left(\frac{{\rm d}}{{\rm d}x} \hat{\psi} (X_j^-)\right) + 
    \theta_0 \push \left(\frac{{\rm d}}{{\rm d}x}\hat{\psi} (X_j^+)\right) \right] 
    + \text{h.c.} \;,
    \label{suppeq:boundary}
\end{align}
where $\hat{\psi}(X_j^+) \equiv \hat{\psi}_{>}(X_j)$, $\hat{\psi}(X_j^-) \equiv \hat{\psi}_{<}(X_j)$, $\theta_0 \coloneqq \theta(0)$, and $\Lambda = \delta(0)/2$ is a formally divergent energy penalty that ensures the wave function is continuous, i.e., $\hat{\psi}(X_j^+) = \hat{\psi}(X_j^-)$. For the numerics, we treat $\Lambda$ as a phenomenological parameter. As $\Lambda$ is increased, the discontinuity in the wave function falls off as $1/\Lambda$, and so does the second line of Eq.~\eqref{suppeq:boundary}. In fact, these other terms do not impose any constraint on the wave function, and we find they also do not alter the numerical convergence to the ground state. Thus, one can drop these without affecting physical predictions, which gives $\hat{K} = \sum_j \hat{K}_j + \Lambda \hat{\Upsilon}_{j,j+1}$, where
\begin{equation}
    \hat{K}_j = \int_{X_{j-1}}^{X_j} \pull\pull {\rm d}x \; \hat{\mathcal{K}}(x)\;, \quad \text{and} \quad
    \hat{\Upsilon}_{j,j+1} \coloneqq 
    \big[ \hat{\psi}(X_j^+) - \hat{\psi}(X_j^-) \big]^{\dagger} 
    \big[ \hat{\psi}(X_j^+) - \hat{\psi}(X_j^-) \big] \;.
    \label{suppeq:KUpsilon}
\end{equation}
Thus, we arrive at the partitioned Hamiltonian $\hat{H} = \sum_j \hat{K}_j + \hat{U}_j + \hat{P}_j + \Lambda \hat{\Upsilon}_{j,j+1}$.

\section{Basis functions and local operators}
\label{suppsec:basis}

As explained in the main text, each segment $j$ is spanned by multiple $n$-body basis functions $\phi^{(j)}_{n,k}(\vec{r})$, with $\vec{r} \equiv \{x_1,x_2,\dots,x_n\}$, which leads to a matrix representation of the local operators. Here we show how to compute these matrix elements in terms of monomial integrals. We also show how to include the cusp constraint which arises from contact interactions.

\subsection{Characterization in terms of monomials}
\label{suppsec:monomials}

Given an $n$-particle basis function $\phi^{(j)}_{n,k}(\vec{r})$, the many-body state may be expressed as
\begin{equation}
    \big| \phi^{(j)}_{n,k} \big\rangle = 
    \int_{X_{j-1}}^{X_j} \pull\pull {\rm d}^n r \; 
    \phi^{(j)}_{n,k}(\vec{r}) \;
    \frac{\hat{\psi}^{\dagger}(x_1) \hat{\psi}^{\dagger}(x_2) \dots \hat{\psi}^{\dagger}(x_n)}{\sqrt{n!}} 
    \push |0\rangle \;,
    \label{suppeq:basisstates}
\end{equation}
where $|0\rangle$ is the vacuum, and the integration limits denote that all coordinates lie in the interval $[X_{j-1}, X_j]$. As we are dealing with bosons, the wave function $\phi^{(j)}_{n,k}(\vec{r})$ is symmetric under exchange of any two coordinates, and the field operators satisfy the commutation $[ \hat{\psi}(x), \hat{\psi}^{\dagger}(x^{\prime}) ] = \delta(x-x^{\prime}) $. The inner product vanishes between states with unequal numbers of particles, while
\begin{equation}
    \big\langle \phi^{(j)}_{n,k} \big| \phi^{(j)}_{n,k^{\prime}} \big\rangle = 
    \int_{X_{j-1}}^{X_j} \pull\pull {\rm d}^n r \push
    \phi^{(j) *}_{n,k}(\vec{r}) \push 
    \phi^{(j)}_{n,k^{\prime}}(\vec{r}) \;.
    \label{suppeq:innerprod}
\end{equation}
For the basis to be orthogonal, (\ref{suppeq:innerprod}) must vanish for $k \neq k^{\prime}$. It is convenient to rescale the coordinates so that the interval runs from $0$ to $1$,
\begin{equation}
    \phi^{(j)}_{n,k}(\vec{r}) \coloneqq 
    \frac{1}{w_j^{n/2}} \push 
    \chi^{(j)}_{n,k} \bigg( \frac{\vec{r} - \vec{R}_{j-1}}{w_j} \bigg) \;.
    \label{suppeq:rescale}
\end{equation}
where $w_j \coloneqq X_j - X_{j-1}$ and $\vec{R}_{j-1}$ is the coordinate $x_1 = x_2 = \dots = x_n = X_{j-1}$. Then Eq.~\eqref{suppeq:innerprod} becomes
\begin{equation}
    \big\langle \phi^{(j)}_{n,k} \big| \phi^{(j)}_{n,k^{\prime}} \big\rangle = 
    \int_0^1 \pull\pull {\rm d}^n r \push
    \chi^{(j) *}_{n,k}(\vec{r}) \push 
    \chi^{(j)}_{n,k^{\prime}}(\vec{r}) \;,
    \label{suppeq:innerprod01}
\end{equation}
so the basis will be orthonormal provided $\chi^{(j)}_{n,k}$ are orthonormal over the interval $[0,1]^n$. 

One can always write $\chi^{(j)}_{n,k} (\vec{r})$ as a sum over symmetrized monomials $\mathbf{p}(\vec{r})$ whose variation in the ``canonical'' sector
\begin{equation}
    \mathcal{S}_n:\;\; 
    0 \leq x_1 \leq x_2 \leq \dots \leq x_n \leq 1
    \label{suppeq:canonical}
\end{equation}
is given by $\mathbf{p}(\vec{r}) = x_1^{p_1} x_2^{p_2} \dots x_n^{p_n}$, where $p_i$ are nonnegative integers. In this sector,
\begin{equation}
    \chi^{(j)}_{n,k}(\vec{r}) = 
    \sum\nolimits_{\mathbf{p}} A^{(j)}_{n,k,\mathbf{p}} \;
    \mathbf{p}(\vec{r}) \;,
    \label{suppeq:monomialexpansion}
\end{equation}
with expansion coefficients $A^{(j)}_{n,k,\mathbf{p}}$, which characterize the basis. In numerical calculations, one has to restrict to a finite basis by constraining $\mathbf{p} \coloneqq \{p_1, p_2, \dots, p_n \}$. For example, the number of monomials of degree $\sum_{i=1}^n p_i = d$ grows as $\binom{d+n-1}{d}$, and we find it convenient to only retain states with $d \leq d_{\text{max}}$. The monomials are not orthogonal, but their inner product $\langle \mathbf{p} | \mathbf{q} \rangle = n! \; \mathcal{I}(\mathbf{p} + \mathbf{q})$ has a simple form
\begin{equation}
    \mathcal{I}(\mathbf{p}) \coloneqq 
    \int_0^1 \pull\pull {\rm d}x_n 
    \int_0^{x_n} \pull\pull {\rm d}x_{n-1} \dots
    \int_0^{x_2} \pull\pull {\rm d}x_1 \;
    x_1^{p_1} x_2^{p_2} \dots x_n^{p_n}
    = \prod_{n^{\prime} = 1}^n \push
    \frac{1}{n^{\prime} + \sum_{i=1}^{n^{\prime}} p_i} \;.
    \label{suppeq:intmon}
\end{equation}
As explained in Sec.~\ref{suppsec:construction}, a convenient way to construct our orthonormal basis is to take the states to be eigenvectors of a Hermitian operator. When expressed in terms of the non-orthogonal monomials, this involves solving a generalized eignevalue problem. Physical constraints, such as the cusp condition arising from short-range interactions, are incorporated by correctly choosing the Hermitian operator.

\subsection{Matrix elements of local operators}
\label{suppsec:localops}

\subsubsection{Field operator}
\label{suppsec:fieldop}

The action of the field operator $\hat{\psi}(x)$ on the basis states in Eq.~\eqref{suppeq:basisstates} can be found by using Bose commutation and the symmetry of $\phi^{(j)}_{n,k}(\vec{r})$ under particle exchange, which gives, for $X_{j-1} \leq x \leq X_j$,
\begin{equation}
    \hat{\psi}(x) \push \big| \phi^{(j)}_{n,k} \big\rangle = 
    \sqrt{n} \int_{X_{j-1}}^{X_j} \pull\pull {\rm d}^{n-1} r \; 
    \phi^{(j)}_{n,k}(x, \vec{r}) \;
    \frac{\hat{\psi}^{\dagger}(x_1) \hat{\psi}^{\dagger}(x_2) \dots \hat{\psi}^{\dagger}(x_{n-1})}{\sqrt{(n-1)!}} 
    \push |0\rangle \;.
    \label{suppeq:psiphi}
\end{equation}
The right-hand side describes a state of $n-1$ particles with the (unnormalized) wave function $\sqrt{n} \push \phi^{(j)}_{n,k}(x,\vec{r})$ -- which is simply the original wave function with one of the positions set to $x$.  Because of the Bose symmetry, it does not matter which particle is selected. 

The nonzero matrix elements of $\hat{\psi}(x)$ are given by [using Eqs.~\eqref{suppeq:innerprod} and \eqref{suppeq:rescale}]
\begin{equation}
    \big\langle \phi^{(j)}_{n-1,k} 
    \big| \push \hat{\psi}(x) \push \big| 
    \phi^{(j)}_{n,k^{\prime}} \big\rangle = 
    \sqrt{n} \int_{X_{j-1}}^{X_j} \pull\pull {\rm d}^{n-1} r \; 
    \phi^{(j) *}_{n-1,k} (\vec{r}) \push 
    \phi^{(j)}_{n,k^{\prime}}(x, \vec{r}) 
    = \sqrt{\frac{n}{w_j}} 
    \int_0^1 \pull\pull {\rm d}^{n-1} r \; 
    \chi^{(j) *}_{n-1,k} (\vec{r}) \push 
    \chi^{(j)}_{n,k^{\prime}}(\tilde{x}, \vec{r}) \;,
\end{equation}
where $\tilde{x} \coloneqq (x - X_{j-1}) / w_j$. As $\chi^{(j)}_{n,k}$ are linear combinations of symmetrized monomials [Eq.~\eqref{suppeq:monomialexpansion}], it suffices to evaluate this integral for any two such monomials, i.e.,
\begin{equation}
    \psi_{\mathbf{p},\mathbf{q}}(\tilde{x}) \coloneqq 
    \sqrt{n} \int_0^1 \pull {\rm d}^{n-1}r \;
    \mathbf{p}(\vec{r}) \;\mathbf{q}(\tilde{x},\vec{r}) \;,
\end{equation}
where $\mathbf{p} = \{p_1, p_2, \dots, p_{n-1}\}$, $\mathbf{q} = \{q_1, q_2, \dots, q_n\}$. Using exchange symmetry, one can write $\psi_{\mathbf{p},\mathbf{q}}(\tilde{x}) = \sqrt{n} (n-1)! \int_{\mathcal{S}_{n-1}}\pull\pull {\rm d}^{n-1}r \;\mathbf{p}(\vec{r}) \;\mathbf{q}(\tilde{x},\vec{r})$ over the canonical sector $\mathcal{S}_{n-1}$ [Eq.~\eqref{suppeq:canonical}]. In this sector, $\mathbf{p}(\vec{r}) = x_1^{p_1} x_2^{p_2} \dots x_{n-1}^{p_{n-1}}$, but the expression for $\mathbf{q}(\tilde{x}, \vec{r})$ depends on the ordering of $\tilde{x}$ relative to the other coordinates. For $\tilde{x} \to 0^+$ (i.e., $x \to X_{j-1}^+$), $q(\tilde{x}, \vec{r}) = \tilde{x}^{q_1} x_1^{q_2} x_2^{q_3} \dots x_{n-1}^{q_n}$, whereas for $\tilde{x} \to 1^-$ ($x \to X_j^-)$, $q(\tilde{x}, \vec{r}) = x_1^{q_1} x_2^{q_2} \dots x_{n-1}^{q_{n-1}} \tilde{x}^{q_n}$. Thus, using Eq.~\eqref{suppeq:intmon},
\begin{align}
    \psi_{\mathbf{p},\mathbf{q}}(0) &= 
    \delta_{q_1,0} \push \sqrt{n} \; (n-1)! \; \mathcal{I}
    \big(\{p_1 + q_2, p_2 + q_3, \dots, p_{n-1} + q_n\}\big) \;,
    \label{suppeq:psipq0}\\[0.1cm]
    \psi_{\mathbf{p},\mathbf{q}}(1) &= 
    \sqrt{n} \; (n-1)! \; \mathcal{I}
    \big(\{p_1 + q_1, p_2 + q_2, \dots, p_{n-1} + q_{n-1}\}\big) \;.
    \label{suppeq:psipq1}
\end{align}
For intermediate values of $\tilde{x}$, the integral can be split into domains where $x_1 \leq \dots \leq x_{n^{\prime}  - 1} \leq \tilde{x} \leq x_{n^{\prime} + 1} \leq \dots \leq x_{n-1}$ for $n^{\prime} = 1, 2, \dots, n$, which gives
\begin{equation}
    \psi_{\mathbf{p}, \mathbf{q}}(\tilde{x}) = 
    \sqrt{n} \; (n-1)! 
    \sum_{n^{\prime} = 1}^n \tilde{x}^{q_{n^{\prime}}} 
    \mathcal{I}_l \big( \{p_1 + q_1, \dots, p_{n^{\prime}-1} + q_{n^{\prime}-1}\}, \tilde{x} \big) \;
    \mathcal{I}_r \big( \tilde{x}, \{p_{n^{\prime}} + q_{n^{\prime} + 1}, \dots, p_{n-1} + q_{n}\} \big) \;,
    \label{suppeq:psipq}
\end{equation}
where 
\begin{align}
    \mathcal{I}_l \big(\{p_1, \dots, p_n\}, \tilde{x}\big) \coloneqq&\; 
    \int_0^{\tilde{x}} \pull\pull {\rm d}x_n 
    \int_0^{x_n} \pull\pull {\rm d}x_{n-1} \dots
    \int_0^{x_2} \pull\pull {\rm d}x_1 \;
    x_1^{p_1} x_2^{p_2} \dots x_n^{p_n} 
    = \tilde{x}^{n + \sum_{i=1}^n p_i} \push
    \mathcal{I} \big(\{p_1, \dots, p_n\}\big) \;,
    \label{suppeq:Il}\\[0.1cm]
    \nonumber
    \mathcal{I}_r \big(\tilde{x}, \{p_1, \dots, p_n\}\big) \coloneqq&\; 
    \int_{\tilde{x}}^1 \pull\pull {\rm d}x_n 
    \int_{\tilde{x}}^{x_n} \pull\pull {\rm d}x_{n-1} \dots
    \int_{\tilde{x}}^{x_2} \pull\pull {\rm d}x_1 \;
    x_1^{p_1} x_2^{p_2} \dots x_n^{p_n} 
    \\
    =&\; \sum_{i=0}^n (-1)^i \;
    \mathcal{I}_l \big(\{p_i, p_{i-1} \dots, p_1\}, \tilde{x}\big) \;
    \mathcal{I} \big(\{p_{i+1}, p_{i+2}, \dots, p_n\}\big)\;,
    \label{suppeq:Ir}
\end{align}
and $\mathcal{I}_l (\{\},\tilde{x}) \coloneqq 1$. Note that $\mathcal{I}_l (\mathbf{p}, 1) = \mathcal{I}_r (0, \mathbf{p}) = \mathcal{I}(\mathbf{p})$. Equation~\eqref{suppeq:psipq} gives $\psi_{\mathbf{p}, \mathbf{q}}(\tilde{x})$ as a polynomial in $\tilde{x}$ of degree $n-1 + q_n + \sum_{i=1}^{n-1} p_i + q_i$. In numerical simulations, we store these polynomial coefficients.

\subsubsection{Density and potential energy}
\label{suppsec:densitypot}

A similar construction applies for the density $\hat{\rho}(x) = \hat{\psi}^{\dagger}(x) \hat{\psi}(x)$. Using Eq.~\eqref{suppeq:psiphi}, one finds the matrix elements
\begin{equation}
    \big\langle \phi^{(j)}_{n,k} 
    \big| \push \hat{\rho}(x) \push \big| 
    \phi^{(j)}_{n,k^{\prime}} \big\rangle = 
    n \int_{X_{j-1}}^{X_j} \pull\pull {\rm d}^{n-1} r \; 
    \phi^{(j) *}_{n,k} (x, \vec{r}) \push 
    \phi^{(j)}_{n,k^{\prime}}(x, \vec{r}) 
    = \frac{n}{w_j} 
    \int_0^1 \pull\pull {\rm d}^{n-1} r \; 
    \chi^{(j) *}_{n,k} (\tilde{x}, \vec{r}) \push 
    \chi^{(j)}_{n,k^{\prime}}(\tilde{x}, \vec{r}) 
    \label{suppeq:rhoelem}
\end{equation}
for $X_{j-1} \leq x \leq X_j$. As before, the integral needs to be calculated only for symmetrized monomials,
\begin{equation}
    \rho_{\mathbf{p},\mathbf{q}}(\tilde{x}) \coloneqq 
    n \int_0^1 \pull {\rm d}^{n-1}r \;
    \mathbf{p}(\tilde{x}, \vec{r}) \; 
    \mathbf{q}(\tilde{x},\vec{r}) \;,
\end{equation}
where $\mathbf{p} = \{p_1, p_2, \dots, p_{n}\}$ and $\mathbf{q} = \{q_1, q_2, \dots, q_n\}$. Using exchange symmetry, $\rho_{\mathbf{p},\mathbf{q}}(\tilde{x}) = n! \int_{\mathcal{S}_{n-1}} \pull\pull {\rm d}^{n-1}r \; \mathbf{p}(\tilde{x}, \vec{r}) \; \mathbf{q}(\tilde{x},\vec{r})$ where $\mathcal{S}_n$ denotes the canonical ordering [Eq.~\eqref{suppeq:canonical}]. At the boundaries $\tilde{x} \to 0^+$ and $\tilde{x} \to 1^-$, both $\mathbf{p}(\tilde{x}, \vec{r})$ and $\mathbf{q}(\tilde{x}, \vec{r})$ reduce to single monomials, yielding [as in Eqs.~\eqref{suppeq:psipq0} and \eqref{suppeq:psipq1}]
\begin{align}
    \rho_{\mathbf{p},\mathbf{q}}(0) &= 
    \delta_{p_1 + q_1,0} \push n! \; \mathcal{I}
    \big(\{p_2 + q_2, p_3 + q_3, \dots, p_n + q_n\}\big) \;,
    \label{suppeq:rhopq0}\\[0.1cm]
    \rho_{\mathbf{p},\mathbf{q}}(1) &= 
    n! \; \mathcal{I}
    \big(\{p_1 + q_1, p_2 + q_2, \dots, p_{n-1} + q_{n-1}\}\big) \;.
    \label{suppeq:rhopq1}
\end{align}
For intermediate $\tilde{x}$, we split the integral into subdomains $x_1 \leq \dots \leq x_{n^{\prime} - 1} \leq \tilde{x} \leq x_{n^{\prime} + 1} \leq \dots \leq x_{n-1}$, obtaining
\begin{equation}
    \rho_{\mathbf{p}, \mathbf{q}}(\tilde{x}) = 
    n! \sum_{n^{\prime} = 1}^n 
    \tilde{x}^{p_{n^{\prime}} + q_{n^{\prime}}} 
    \mathcal{I}_l \big( \{p_1 + q_1, \dots, p_{n^{\prime}-1} + q_{n^{\prime}-1}\}, \tilde{x} \big) \;
    \mathcal{I}_r \big( \tilde{x}, \{p_{n^{\prime} + 1} + q_{n^{\prime} + 1}, \dots, p_n + q_{n}\} \big) \;,
    \label{suppeq:rhopq}
\end{equation}
where $\mathcal{I}_l$ and $\mathcal{I}_r$ are given by Eqs.~\eqref{suppeq:Il} and \eqref{suppeq:Ir}. Again, $\rho_{\mathbf{p}, \mathbf{q}} (\tilde{x})$ is a polynomial in $\tilde{x}$ of degree $n-1 + \sum_{i=1}^{n} p_i + q_i$, and we store the coefficients.

The matrix elements of the potential energy $\hat{P}_j = \int_{X_{j-1}}^{X_j} \pull\pull {\rm d}x \push V(x) \push \hat{\rho}(x)$ can be obtained from those of the density in Eq.~\eqref{suppeq:rhoelem}. In particular, for two symmetrized monomials, one calculates
\begin{equation}
    V^{(j)}_{\mathbf{p}, \mathbf{q}} \coloneqq 
    \int_0^1 \pull\pull {\rm d}{\tilde{x}} \;
    V(X_{j-1} + w_j \tilde{x}) \; 
    \rho_{\mathbf{p}, \mathbf{q}}(\tilde{x}) \;,
    \label{suppeq:Vjpq}
\end{equation}
which reduces to computing moments of $V(x)$, since $\rho_{\mathbf{p}, \mathbf{q}}(\tilde{x})$ is a polynomial in $\tilde{x}$. For a sinusoidal potential $V(x) = V_0 \cos^2 k x$, these moments can be found exactly using
\begin{equation}
    \int_0^1 \pull\pull {\rm d}\tilde{x} \push 
    \tilde{x}^{p-1} e^{{\rm i} \tilde{k} \tilde{x}} = 
    ({\rm i}/k)^p \gamma(p, -{\rm i}k)
\end{equation}
$\forall p \geq 1$, where $\gamma(p,z)$ is the lower incomplete gamma function \cite{zhang1996computation}, distinct from the interaction strength $\gamma$.

\subsubsection{Kinetic energy}
\label{suppsec:kinetic}

The kinetic energy within the $j$-th segment is given by $\hat{K}_j  = (1/2) \int_{X_{j-1}}^{X_j} \pull\pull {\rm d}x \push [\partial_x \hat{\psi}^{\dagger}(x)] \push [\partial_x \hat{\psi}(x)]$. To find its matrix elements, we use Eq.~\eqref{suppeq:psiphi} and the exchange symmetry of the basis functions, obtaining
\begin{equation}
    \big\langle \phi^{(j)}_{n,k} 
    \big| \push \hat{K}_j \push \big| 
    \phi^{(j)}_{n,k^{\prime}} \big\rangle = 
    \frac{1}{2} \int_{X_{j-1}}^{X_j} \pull\pull {\rm d}^n r \; 
    \vec{\nabla} \phi^{(j) *}_{n,k} (\vec{r}). 
    \vec{\nabla} \phi^{(j)}_{n,k^{\prime}}(\vec{r}) 
    = \frac{1}{2 w_j^2} 
    \int_0^1 \pull\pull {\rm d}^n r \; 
    \vec{\nabla} \chi^{(j) *}_{n,k} (\vec{r}). 
    \vec{\nabla} \chi^{(j)}_{n,k^{\prime}}(\vec{r}) \;.
    \label{suppeq:Kelem}
\end{equation}
Replacing $\chi^{(j)}_{n,k}(\vec{r})$ with a symmetrized monomial [Eq.~\eqref{suppeq:monomialexpansion}], we only need to evaluate
\begin{equation}
    K_{\mathbf{p}, \mathbf{q}} \coloneqq 
    \frac{1}{2} \int_0^1 \pull\pull {\rm d}^n r \; 
    \vec{\nabla} \mathbf{p} (\vec{r}). 
    \vec{\nabla} \mathbf{q} (\vec{r}) =
    \frac{n!}{2} \int_{\mathcal{S}_n} \pull\pull {\rm d}^n r \; 
    \vec{\nabla} \mathbf{p} (\vec{r}). 
    \vec{\nabla} \mathbf{q} (\vec{r}) \;,
\end{equation}
where $\mathbf{p} = \{p_1, p_2, \dots, p_{n}\}$ and $\mathbf{q} = \{q_1, q_2, \dots, q_n\}$. In sector $\mathcal{S}_n$, $\partial_{x_i} \mathbf{p}(\vec{r}) = p_i x_i^{p_i - 1} \prod_{i^{\prime} \neq i} x_{i^{\prime}}^{p_{i^{\prime}}}$. Thus, Eq.~\eqref{suppeq:intmon} yields
\begin{equation}
    K_{\mathbf{p}, \mathbf{q}} = 
    \frac{n!}{2} \push \sum_{i=1}^n \push p_i q_i \; \mathcal{I} \big( \texttt{incr}(\mathbf{p} + \mathbf{q}, i, -2) \big) \;,
\end{equation}
where $\texttt{incr}(\mathbf{p}, i, s) \coloneqq \{p_1, \dots, p_{i-1}, p_i + s, p_{i+1}, \dots, p_n\}$, i.e., it increments the $i$-th element by $s$. Note that one can also extract matrix elements of the kinetic energy density $\hat{\mathcal{K}}(x)$ in Eq.~\eqref{suppeq:K} using the procedure in Sec.~\ref{suppsec:densitypot}.

\subsubsection{Interaction energy}
\label{suppsec:interaction}

For the interaction energy $\hat{U}_j$ in Eq.~\eqref{suppeq:UjandPj}, we again use the action of the field operator in Eq.~\eqref{suppeq:psiphi} to obtain
\begin{align}
    \big\langle \phi^{(j)}_{n,k} 
    \big| \push \hat{U}_j \push \big| 
    \phi^{(j)}_{n,k^{\prime}} \big\rangle &= 
    g \binom{n}{2} \int_{X_{j-1}}^{X_j} \pull\pull {\rm d}x 
    \int_{X_{j-1}}^{X_j} \pull\pull {\rm d}^{n-2} r \; 
    \phi^{(j) *}_{n,k} (x, x, \vec{r}) \;
    \phi^{(j)}_{n,k^{\prime}}(x, x, \vec{r}) 
    \\[0.2cm]
    &= g \int_{X_{j-1}}^{X_j} \pull\pull {\rm d}^n r 
    \sum_{i < i^{\prime}} \delta(x_i - x_{i^{\prime}}) \;
    \phi^{(j) *}_{n,k} (\vec{r}) \;
    \phi^{(j)}_{n,k^{\prime}}(\vec{r}) 
    = \frac{g}{w_j} \int_0^1 \pull\pull {\rm d}^n r \push
    \sum_{i < i^{\prime}} \delta(x_i - x_{i^{\prime}}) \;
    \chi^{(j) *}_{n,k} (\vec{r}) \;
    \chi^{(j)}_{n,k^{\prime}}(\vec{r}) \;.
\end{align}
Therefore, constructing the matrix for $\hat{U}_j$ boils down to evaluating
\begin{equation}
    U_{\mathbf{p}, \mathbf{q}} \coloneqq 
    \int_0^1 \pull\pull {\rm d}^n r \push
    \sum_{i < i^{\prime}} \delta(x_i - x_{i^{\prime}}) \;
    \mathbf{p} (\vec{r}) \;
    \mathbf{q} (\vec{r}) 
    = \frac{n!}{2} \int_{\mathcal{S}_n} \pull\pull {\rm d}^n r \push 
    \sum_{i=1}^{n-1} \delta(x_i - x_{i+1}) \;
    \mathbf{p} (\vec{r}) \;
    \mathbf{q} (\vec{r}) 
\end{equation}
for symmetrized monomials $\mathbf{p} = \{p_1, p_2, \dots, p_{n}\}$ and $\mathbf{q} = \{q_1, q_2, \dots, q_n\}$. Note, in sector $\mathcal{S}_n$, $x_1 \leq x_2 \leq \dots \leq x_n$, so we have the delta function only between neighboring coordinates $x_i$ and $x_{i+1}$. Substituting the monomial expressions for $\mathbf{p}(\vec{r})$ and $\mathbf{q}(\vec{r})$, and using Eq.~\eqref{suppeq:intmon}, we find
\begin{equation}
    U_{\mathbf{p}, \mathbf{q}} =  
    \frac{n!}{2} \push \sum_{i=1}^{n-1} \; \mathcal{I} \big( \texttt{merge}(\mathbf{p} + \mathbf{q}, i) \big) \;,
\end{equation}
where $\texttt{merge}(\mathbf{p},i) \coloneqq \{p_1, \dots, p_{i-1}, p_i + p_{i+1}, p_{i+2}, \dots, p_n \}$, i.e., it merges the $i$- and $i+1$-th elements. As with $\hat{\mathcal{K}}(x)$, the matrix elements of the interaction energy density $\hat{\mathcal{U}}(x)$ in Eq.~\eqref{suppeq:U} can be obtained using the method in Sec.~\ref{suppsec:densitypot}.

\subsection{Basis construction for contact interactions}
\label{suppsec:construction}

The ideal choice of basis functions, $\chi^{(j)}_{n,k}(\vec{r})$, would have three properties: (i) A small number of these states should accurately approximate the low-energy eigenstates of the Hamiltonian, (ii) these states should smoothly connect to the wave functions in neighboring sectors, and (iii) they should be orthogonal to one another. The latter can be ensured by taking them to be eigenstates of a Hermitian operator -- and choosing the basis is equivalent to choosing the operator.  
 
Insight into the choice of operator comes from the one-particle sector, where selecting the basis functions is related to deciding on a functional form for splines which will be used to piecewise describe the ground state of the Schr\"{o}dinger equation. In that case, one might naively choose the single-particle basis functions on $[0,1]$ to be solutions to Laplace's equation, $\partial_{x_1}^2 \chi=\lambda \chi$. Depending on boundary conditions, $\chi=\cos(\pi kx_1)$ with $k=0,1,\ldots$ or $\chi=\sin(\pi k x_1)$ with $k=1,2,\ldots$. Neither set of states is amenable to continuously connecting across segments -- as either the basis function or its derivative vanishes at the boundaries.  The solution is to modify the operator so that $x_1=0$ and $x_1=1$ are regular singular points; for example choosing them to be solutions to Legendre's equation: $\partial_{x_1} [x_1(1-x_1)\partial_{x_1} \chi]=\lambda \chi$.  The solutions are Legendre polynomials $P_k(2x_1-1)$ \cite{zhang1996computation}, and the resulting wave function expansion amounts to using polynomial splines.  Colloquially, one imagines that the factor $x_1(1-x_1)$ ``absorbs" the boundary conditions, allowing basis functions to have both nonzero amplitude and slope at the segment boundaries. An equivalent construction of the Legendre polynomials is to perform a Gram-Schmidt orthogonalization on the monomials $\{1, x_1, x_1^2, \dots, x_1^k\}$.
 
In our problem the many-body wave function has a kink whenever two particles coincide. Thus, the expansion will perform better if the basis functions also have this kink, motivating the modified Legendre equation,
\begin{equation}
    - \frac{1}{2} \sum_{i=1}^n \frac{\partial}{\partial x_i} 
    \bigg[ x_i (1 - x_i) \push \frac{\partial}{\partial x_i} \push \chi^{(j)}_{n,k}(\vec{r}) \bigg] +
    c_j \sum_{i < i^{\prime}} x_i (1 - x_i) \push 
    \delta(x_i - x_{i^{\prime}}) \; \chi^{(j)}_{n,k}(\vec{r}) = 
    \mathcal{E}^{(j)}_{n,k} \; \chi^{(j)}_{n,k}(\vec{r}) \;,
    \label{suppeq:basisgen}
\end{equation}
where $c_j$ gives the slope discontinuity, $\partial_{x_i}  \chi^{(j)}_{n,k} (x_i \to x_{i^{\prime}}^+) - \partial_{x_i} \chi^{(j)}_{n,k} (x_i \to x_{i^{\prime}}^-) = c_j \push \chi^{(j)}_{n,k}(x_i = x_{i^{\prime}})$. As the coordinates of $\chi$ are stretched by a factor of $w_j$ relative to the physical coordinates [Eq.~\eqref{suppeq:rescale}], we require $c_j = w_j g$. For uniform segments, $w_j = \text{constant}$, thus $c_j$ and $\chi^{(j)}_{n,k}$ do not depend on $j$.

We expand Eq.~\eqref{suppeq:basisgen} on the (non-orthogonal) symmetrized monomials. Defining 
\begin{eqnarray}
\hat{K}_{\mathcal{L}}[\chi(\vec{r})]&\coloneqq&- \frac{1}{2} \sum_{i=1}^n \frac{\partial}{\partial x_i} 
    \bigg[ x_i (1 - x_i) \push \frac{\partial}{\partial x_i} \push \chi(\vec{r}) \bigg]\\[0.1cm]
\hat{U}_{\mathcal{L}}[\chi(\vec{r})]&\coloneqq&
 \sum_{i < i^{\prime}} x_i (1 - x_i) \push 
    \delta(x_i - x_{i^{\prime}}) \; \chi(\vec{r}) 
\end{eqnarray}
and their sum $\hat H^{(j)}_{\mathcal{L}}=
\hat{K}_{\mathcal{L}}+c_j \hat{U}_{\mathcal{L}}$, the matrix elements are
\begin{align}
    \big\langle \mathbf{p} \big| \hat{K}_{\mathcal{L}} \big| \mathbf{q} \big\rangle &= 
    \frac{n!}{2} \int_{\mathcal{S}_n} \pull\pull {\rm d}^n r 
    \sum_{i=1}^n x_i (1 - x_i) \push 
    [\partial_{x_i} \mathbf{p} (\vec{r})] \push 
    [\partial_{x_i} \mathbf{q} (\vec{r})] \;,
    \label{suppeq:KL}\\[0.1cm]
    \big\langle \mathbf{p} \big| \hat{U}_{\mathcal{L}} \big| \mathbf{q} \big\rangle &= 
    \frac{n!}{2} \int_{\mathcal{S}_n} \pull\pull {\rm d}^n r \push 
    \sum_{i=1}^{n-1} x_i (1 - x_i) \push 
    \delta(x_i - x_{i+1}) \push
    \mathbf{p} (\vec{r}) \push
    \mathbf{q} (\vec{r}) \;.
    \label{suppeq:UL}
\end{align}
As before, $\mathbf{p} = \{p_1, p_2, \dots, p_{n}\}$, $\mathbf{q} = \{q_1, q_2, \dots, q_n\}$, and $\mathcal{S}_n$ stands for the canonical ordering in Eq.~\eqref{suppeq:canonical}. Substituting $\mathbf{p}(\vec{r}) = x_1^{p_1} x_2^{p_2} \dots x_n^{p_n}$ and $\mathbf{q}(\vec{r}) = x_1^{q_1} x_2^{q_2} \dots x_n^{q_n}$, we find, similar to Secs.~\ref{suppsec:kinetic} and \ref{suppsec:interaction},
\begin{align}
    \big\langle \mathbf{p} \big| \hat{K}_{\mathcal{L}} \big| \mathbf{q} \big\rangle =&\; 
    \frac{n!}{2} \push \sum_{i=1}^n \push p_i q_i 
    \Big[ \push \mathcal{I} \big( \texttt{incr}(\mathbf{p} + \mathbf{q}, i, -1) \big) 
    - \mathcal{I}(\mathbf{p} + \mathbf{q}) \Big] \;,
    \label{suppeq:KLpq}\\[0.1cm]
    \big\langle \mathbf{p} \big| \hat{U}_{\mathcal{L}} \big| \mathbf{q} \big\rangle =&\; 
    \frac{n!}{2} \push \sum_{i=1}^{n-1} \Big[ \push 
    \mathcal{I} \Big( \texttt{incr} \big( \texttt{merge}(\mathbf{p} + \mathbf{q},i), i, 1 \big) \Big) 
    - \mathcal{I} \Big( \texttt{incr} \big( \texttt{merge}(\mathbf{p} + \mathbf{q},i), i, 2 \big) \Big) \Big] \;,
    \label{suppeq:ULpq}
\end{align}
where $\mathcal{I}(\mathbf{p})$ is given by Eq.~\eqref{suppeq:intmon}. These expressions reduce Eq.~\eqref{suppeq:basisgen} to a generalized eigenvalue problem $H^{(j)}_{\mathcal{L}} A^{(j)}_{n,k} = \smash{\mathcal{E}^{(j)}_{n,k}} \push \smash{O_n A^{(j)}_{n,k}}$, where $\smash{A^{(j)}_{n,k}}$ are the expansion coefficients of $\smash{\chi^{(j)}_{n,k}}$ in terms of the monomials [Eq.~\eqref{suppeq:monomialexpansion}] and $O_n$ is the matrix of inner products between the monomials, given by $\langle \mathbf{p} | \mathbf{q} \rangle = n! \; \mathcal{I}(\mathbf{p} + \mathbf{q})$. The energy spectrum $\mathcal{E}^{(j)}_{n,k}$ provides a natural ordering of the basis states, which can be truncated at high energies. Although one is solving a many-body problem in generating the basis, the complexity is greatly reduced compared to the original problem, as the number of monomials is limited if the number of particles in a segment, $n$, is sufficiently small.

\section{Simulation parameters}
\label{suppsec:simul}

As outlined above, we construct the basis by solving an eigenvalue problem, Eq.~\eqref{suppeq:basisgen}, in the space of symmetrized monomials of maximum degree $d_{\text{max}}$. The number of monomials grows as $N_{\text{mon}} = \binom{n + d_{\text{max}}}{n}$. Provided the segments are sufficiently narrow, or the repulsive interactions are sufficiently strong ($\gamma \gg 1$), we retain only a few or no basis states for larger $n$. Table~\ref{supptab:basis5} enumerates the total number of monomials and the number of basis states we keep in a typical calculation with strong interactions. Figure~\ref{suppfig:occupations} shows the average weight, $\bar{N}_{n,k}$, of each basis function in the ground state.  These are calculated by averaging over the reduced density matrices describing individual segments, and $\sum_{n,k} \bar{N}_{n,k}=1$. The basis is ordered so that larger $k$ corresponds to larger ${\cal E}_{n,k}$. The weights fall off strongly with both $n$ and $k$, justifying our truncation.  The $k$ dependence is well approximated by a power law, and roughly the same power law is seen for each $n$.

\begin{table}[!htb]
\begin{minipage}[!htb]{0.4\linewidth}
\centering
    \caption{\label{supptab:basis5}Number of monomials and basis states we keep in each segment for the calculation in Fig.~\ref{fig:samplecalc} of the main text.  Here,  $d_{\text{max}} = 4$, $N=5$, $M=8$, and $\gamma=50$.
    }
    \begin{ruledtabular}
    \def\arraystretch{1.1}
    \begin{tabular}{R R R}
         n & N_{\text{mon}} & N_{\text{basis}}
         \vspace{0.05cm}\\ \hline 
         0 & 1 & 1 \\
         1 & 5 & 5 \\
         2 & 15 & 15 \\
         3 & 35 & 35 \\
         4 & 70 & 10 \\
         5 & 126 & 2 \\ \hline
         \text{Total} & 252 & 68
    \end{tabular}
    \end{ruledtabular}
\end{minipage}\hspace{1cm}
\begin{minipage}[!htb]{0.5\linewidth}
\centering
\includegraphics[width=\textwidth]{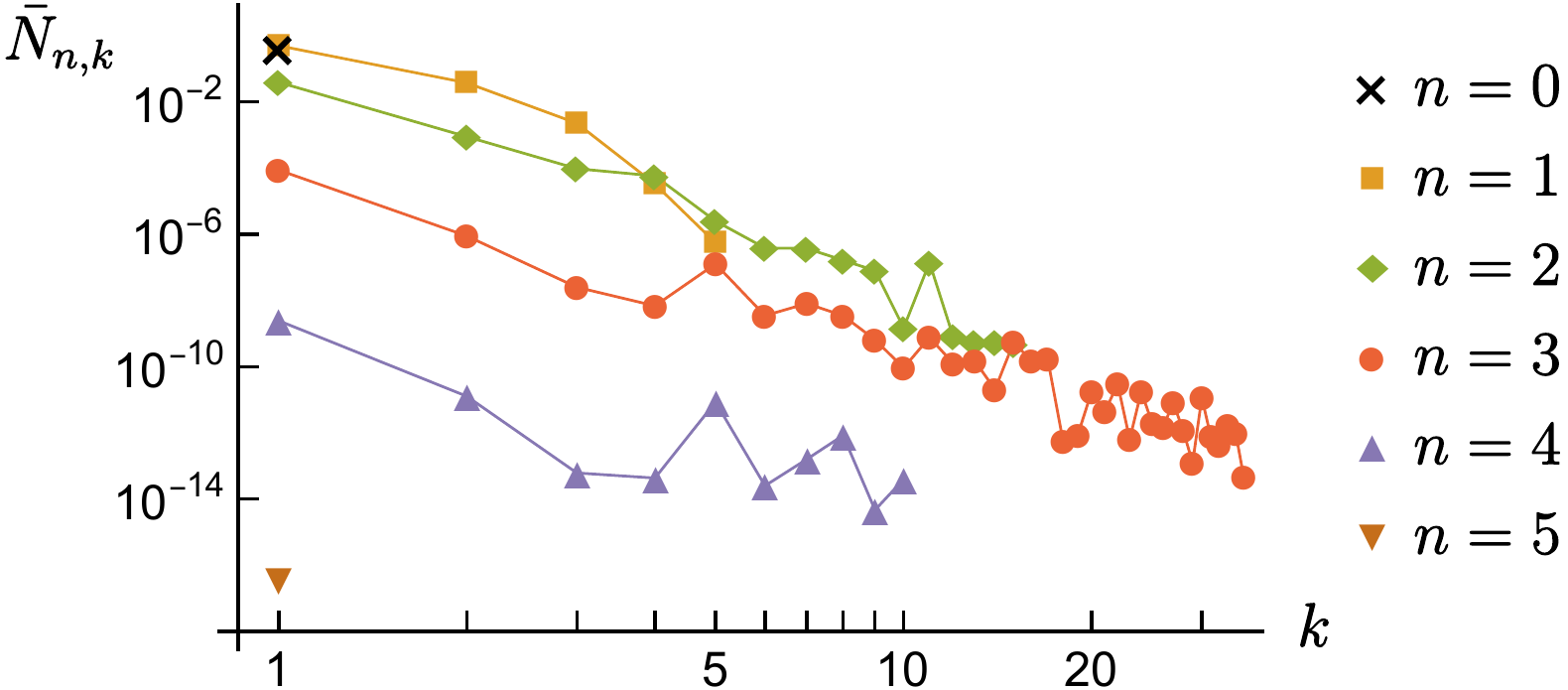}
\captionof{figure}{Average occupation of the basis states enumerated in Table~\ref{supptab:basis5} in the ground state shown in Fig.~\ref{fig:samplecalc}. The basis states are indexed by their eigenvalue $\mathcal{E}_{n,k}$ in Eq.~\eqref{suppeq:basisgen}.
}
\label{suppfig:occupations}
\end{minipage}
\end{table}

Once the basis is generated, one can represent the local operators as matrices following Sec.~\ref{suppsec:localops} and proceed to the DMRG sweeps, for which we used version 3.0.0 of the ITensor library in C++ \cite{supp_itensor}. As discussed in the main text, we run multiple DMRG cycles where the energy penalty $\Lambda$ is successively increased in powers of 10. Here, the practical objective is to produce a good initial state for the final cycle as quickly as possible. Thus, we start from a small maximum bond dimension $D_{\text{max}}$ and gradually increase it with $\Lambda$ to speed up the initial cycles, as shown in Table~\ref{supptab:sweepparams5}. Throughout, we discard singular values below a truncation cutoff $\epsilon_{\text{trunc}} = 10^{-14}$. For the final few cycles, $D_{\text{max}}$ is sufficiently large that this threshold is exceeded on all bonds. For each value of $\Lambda$, we sweep back and forth until the relative change in the total energy (including the discontinuity penalty) between consecutive sweeps is less than a convergence threshold  $\epsilon_{\text{conv}}$, which we lower with increasing $\Lambda$ (as in Table~\ref{supptab:sweepparams5}). We use a Davidson eigensolver with up to a few tens of maximum iterations $\nu_{\text{max}}$, for which our thresholds are typically met after a small number of sweeps. We terminate the program if, at the end of a cycle, the discontinuity $\sum_j \langle \hat{\Upsilon}_{j,j+1} \rangle / L$ has dropped below a target $T_{\text{disc}} = 10^{-12}$. The CPU- and wall times are measured in seconds for each cycle and for the entire DMRG program using the \texttt{clock()} and \texttt{chrono::high\_resolution\_clock::now()} functions, respectively, in C++ on Linux desktops. Since we used quad-core CPUs, wall times are about a quarter of the CPU times.

\begin{table}[!htb]
    \centering
    \caption{\label{supptab:sweepparams5}DMRG parameters for successive cycles with increasing penalty $\Lambda$ for the system in Table~\ref{supptab:basis5} and Fig.~\ref{fig:samplecalc}: $\nu_{\text{max}}$ is the maximum eigensolver iteration, $\epsilon_{\text{trunc}}$ is the singular-value cutoff, $D_{\text{max}}$ is the maximum bond dimension, $\epsilon_{\text{conv}}$ is the convergence threshold, and $D$, $N_{\text{sw}}$, $\Delta t_{\text{CPU}}$, $\Delta t_{\text{wall}}$ are the resulting bond dimension, number of sweeps, CPU- and wall times.
    }
    \begin{ruledtabular}
    \def\arraystretch{1.25}
    \begin{tabular}{R R R R R | R R R R}
         \Lambda L & \nu_{\text{max}} & \epsilon_{\text{trunc}} & D_{\text{max}} & \epsilon_{\text{conv}} & D & N_{\text{sw}} & \Delta t_{\text{CPU}}(s) & \Delta t_{\text{wall}}(s) 
         \vspace{0.05cm}\\ \hline 
         10^1 & 30 & 10^{-14} & 20 & 10^{-3} & 20 & 3 & 45 & 12 \\
         10^2 & 40 & 10^{-14} & 30 & 10^{-4} & 30 & 3 & 109 & 28 \\
         10^3 & 40 & 10^{-14} & 40 & 10^{-5} & 40 & 4 & 203 & 51 \\
         10^4 & 40 & 10^{-14} & 50 & 10^{-6} & 50 & 4 & 257 & 64 \\
         10^5 & 30 & 10^{-14} & 100 & 10^{-7} & 58 & 5 & 255 & 64 \\
         10^6 & 20 & 10^{-14} & 200 & 10^{-8} & 48 & 5 & 112 & 28 
    \end{tabular}
    \end{ruledtabular}
\end{table}

\begin{figure}[!htb]
	\includegraphics[width=1\linewidth]{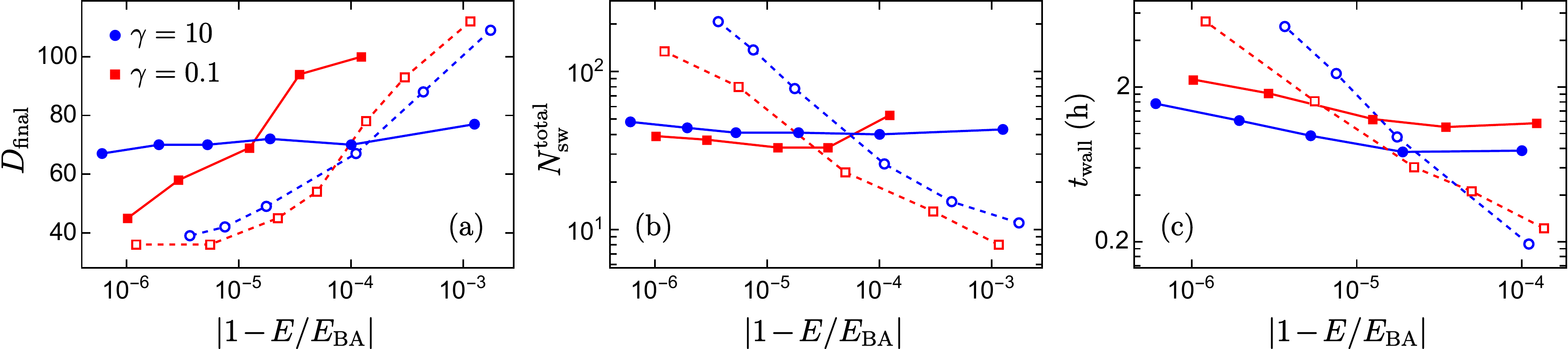}
	\centering
	\caption{(a) Final bond dimension, (b) total number of sweeps, and (c) total wall time as a function of the relative error in the ground-state energy for $N=10$ using cDMRG (solid lines) and the discretized model (dashed lines), corresponding to Fig.~\ref{fig:benchmarking}. The basis and DMRG parameters are given in Tables~\ref{supptab:basis10}--\ref{supptab:sweepparams10B}. Dashed lines in (b) approximately follow $N_{\text{sw}}^{\text{total}} \sim |1-E/E_{\text{BA}}|^{-1/2}$.
	}
	\label{suppfig:benchmarking}
\end{figure}

Table~\ref{supptab:basis10} shows what basis states were used for benchmarking against discretization in Fig.~\ref{fig:benchmarking} of the main text. The corresponding DMRG parameters are listed in Tables~\ref{supptab:sweepparams10A} and \ref{supptab:sweepparams10B}. For the discretized model in Eq.~\eqref{eq:hamildisc}, we employed a standard DMRG cycle with $\epsilon_{\text{trunc}} = 10^{-14}$ and $\epsilon_{\text{conv}} = 10^{-8}$, same as in the final cycle of cDMRG; we used $\nu_{\text{max}} = 3$ which produced good convergence, and although $D_{\text{max}}$ was set to 1000, the actual bond dimensions were comparable to those found using cDMRG, as shown in Fig.~\ref{suppfig:benchmarking}(a). We varied the number of segments and grid points to control the error $\varepsilon$ in the ground-state energy. Figure~\ref{suppfig:benchmarking}(b) shows that the total number of sweeps is relatively independent of $\varepsilon$ for cDMRG, but scales roughly as $\varepsilon^{-1/2}$ for discretization. The wall time for the entire DMRG algorithm, plotted in Fig.~\ref{suppfig:benchmarking}(c), exhibits a similar scaling as the corresponding CPU time in Fig.~\ref{fig:benchmarking}(b) of the main text.

\begin{table}[!htb]
    \centering
    \caption{\label{supptab:basis10}Number of retained basis states, $N_{\text{basis}}$, for $N=10$ in three cases: A and B were used in benchmarking for $\gamma=10$ and $\gamma=0.1$, respectively, in Fig.~\ref{fig:benchmarking}. C was used in Figs.~\ref{fig:inhomogeneous} and \ref{fig:entanglement} to explore ground states in a sinusoidal potential with $M=20$.
    }
    \begin{ruledtabular}
    \def\arraystretch{1.3}
    \begin{tabular}{L | R | R R R R R R R R R R | R}
         \text{Label} & d_{\text{max}} & n=1 & n=2 & n=3 & n=4 & n=5 & n=6 & n=7 & n=8 & n=9 & n=10 & \text{Total}
         \vspace{0.05cm}\\ \hline 
         \text{A} & 3 & 4 & 10 & 20 & 35 & 50 & 30 & 15 & 5 & 2 & 0 & 172 \\
         \text{B} & 3 & 4 & 10 & 20 & 35 & 56 & 70 & 60 & 50 & 35 & 20 & 361 \\
         \text{C} & 4 & 5 & 15 & 35 & 70 & 90 & 50 & 25 & 10 & 5 & 0 & 306
    \end{tabular}
    \end{ruledtabular}
\end{table}

\begin{table}[!htb]
\begin{minipage}{0.47\linewidth}
\centering
    \caption{\label{supptab:sweepparams10A}DMRG parameters used in benchmarking for $N=10$ and $\gamma=10$ in Fig.~\ref{fig:benchmarking}. See Table~\ref{supptab:sweepparams5} for notation.
    }
    \begin{ruledtabular}
    \def\arraystretch{1.25}
    \begin{tabular}{R R R R R}
         \Lambda L & \nu_{\text{max}} & \epsilon_{\text{trunc}} & D_{\text{max}} & \epsilon_{\text{conv}}
         \vspace{0.05cm}\\ \hline 
         10^1 & 20 & 10^{-14} & 20 & 10^{-4} \\
         10^2 & 30 & 10^{-14} & 30 & 10^{-5} \\
         10^3 & 30 & 10^{-14} & 40 & 10^{-6} \\
         10^4 & 30 & 10^{-14} & 50 & 10^{-7} \\
         10^5 & 20 & 10^{-14} & 70 & 10^{-8} \\
         10^6 & 10 & 10^{-14} & 120 & 10^{-8}
    \end{tabular}
    \end{ruledtabular}
\end{minipage}\hspace{0.05\textwidth}
\begin{minipage}{0.47\linewidth}
\centering
    \caption{\label{supptab:sweepparams10B}DMRG parameters used for $N=10$ and $\gamma=0.1$ in Fig.~\ref{fig:benchmarking} and in the presence of a potential in Figs.~\ref{fig:inhomogeneous} and \ref{fig:entanglement}.
    }
    \begin{ruledtabular}
    \def\arraystretch{1.25}
    \begin{tabular}{R R R R R}
         \Lambda L & \nu_{\text{max}} & \epsilon_{\text{trunc}} & D_{\text{max}} & \epsilon_{\text{conv}}
         \vspace{0.05cm}\\ \hline 
         10^1 & 30 & 10^{-14} & 10 & 10^{-4} \\
         10^2 & 40 & 10^{-14} & 20 & 10^{-5} \\
         10^3 & 40 & 10^{-14} & 30 & 10^{-6} \\
         10^4 & 40 & 10^{-14} & 40 & 10^{-7} \\
         10^5 & 30 & 10^{-14} & 60 & 10^{-7} \\
         10^6 & 20 & 10^{-14} & 100 & 10^{-8}
    \end{tabular}
    \end{ruledtabular}
\end{minipage}
\end{table}

\begin{figure}[!htb]
	\includegraphics[width=1\linewidth]{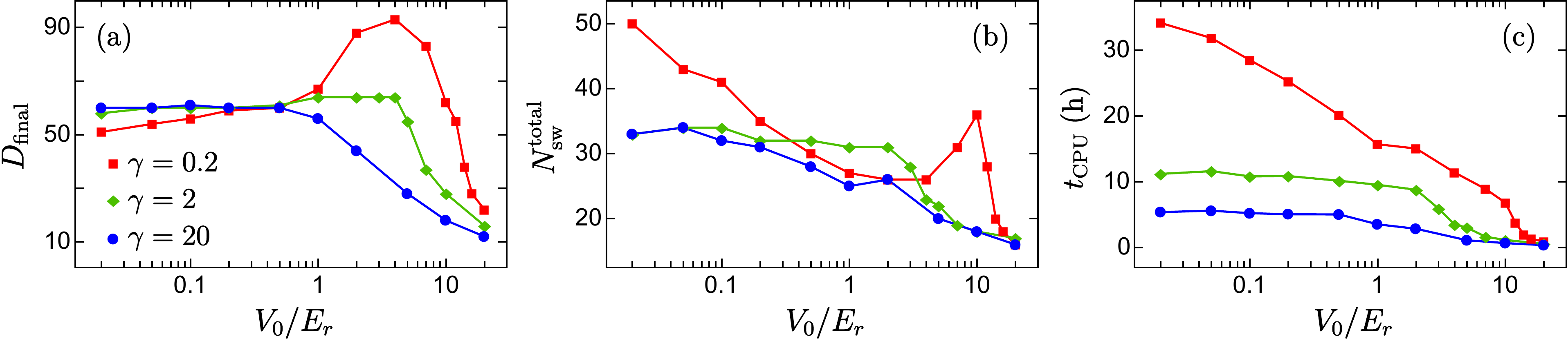}
	\centering
	\caption{(a) Final bond dimension, (b) total number of sweeps, and (c) total CPU time as a function of the potential depth $V_0$, corresponding to Figs.~\ref{fig:inhomogeneous} and \ref{fig:entanglement}, with $N=10$, $M=20$, basis C in Table~\ref{supptab:basis10} and DMRG parameters in Table~\ref{supptab:sweepparams10B}.
	}
	\label{suppfig:potential}
\end{figure}

The ground states in the presence of a sinusoidal potential were obtained using basis C in Table~\ref{supptab:basis10} and the sweep parameters in Table~\ref{supptab:sweepparams10B}. The resulting bond dimensions, sweep numbers, and CPU times are shown in Fig.~\ref{suppfig:potential}. For weak interactions, the first two exhibit peaks where the ground state changes from a Mott insulator to a superfluid. As expected, the CPU time is maximum at weak interactions and weak potentials where the ground state is the most delocalized, requiring a large number of sweeps to populate all basis states.

\section{Splitting a basis into finer segments}
\label{suppsec:split}

As discussed in the main text, one needs to split a basis into finer segments for multiscale DMRG approaches \cite{supp_dolfi2012multigrid} and for obtaining the spatial entanglement at any point other than the segment boundaries. It suffices to consider a single segment with rescaled basis functions $\chi_{n,k}(\vec{r})$ defined over $[0,1]^n$ [see Eq.~\eqref{suppeq:rescale}]. For notational simplicity, we drop the segment label $j$ in this section. To split $\chi_{n,k}$ at at intermediate point $\tilde{x}$, we first construct basis functions $\chi^{\pm}_{n,k}(\vec{r})$ over the left and right zones, $[0,\tilde{x}]^n$ and $[\tilde{x},1]^n$, in terms of symmetrized monomials, as described below. Then the task is to decompose $\chi_{n,k}(\vec{r})$ in the tensor product basis $\chi^+ \otimes \chi^-$, i.e., 
\begin{equation}
    \chi_{n,k}(\vec{r}) = 
    \sum_{n^{\prime}=0}^n \; \sum_{k^+, k^-} 
    \mathcal{C}^{n,k}_{n^{\prime}, k^+, k^-}\; \mathcal{S} \big[
    \chi^+_{n^{\prime}, k^+}(x_1, x_2, \dots, x_{n^{\prime}})\;  
    \chi^-_{n-n^{\prime}, k^-}(x_{n^{\prime}+1}, x_{n^{\prime}+2}, \dots, x_n) \big] \;,
\end{equation}
where $\mathcal{S}$ symmetrizes all $n$ coordinates and the transformation coefficients $\mathcal{C}^{n,k}_{n^{\prime}, k^+, k^-}$ characterize the splitting. Using orthonormality and exchange symmetry of the basis functions, one finds
\begin{flalign}
    \mathcal{C}^{n,k}_{n^{\prime}, k^+, k^-} \pull &= 
    \int_0^1 \pull\pull {\rm d}^n r \; 
    \chi_{n,k}(\vec{r}) \; \mathcal{S} \big[
    \chi^{+ *}_{n^{\prime}, k^+}(x_1, x_2, \dots, x_{n^{\prime}})\;  
    \chi^{- *}_{n-n^{\prime}, k^-}(x_{n^{\prime}+1}, x_{n^{\prime}+2}, \dots, x_n) \big] 
    \\[0.1cm]
    &= \binom{n}{n^{\prime}}^{\pull\pull 1/2} \pull\pull 
    \int_0^{\tilde{x}} \pull\pull {\rm d}x_1 \dots\pull \int_0^{\tilde{x}} \pull\pull {\rm d}x_{n^{\prime}} 
    \int_{\tilde{x}}^1 \pull\pull {\rm d}x_{n^{\prime}+1} \; \dots\pull \int_{\tilde{x}}^1 \pull\pull {\rm d}x_{n} \;
    \chi_{n,k}(\vec{r}) \; 
    \chi^{+ *}_{n^{\prime}, k^+}(x_1, \dots, x_{n^{\prime}}) \; 
    \chi^{- *}_{n-n^{\prime}, k^-}(x_{n^{\prime}+1}, \dots, x_n) \;. \hspace{-0.3cm} &
\end{flalign}
As the basis functions are given in terms of symmetrized monomials [see Eq.~\eqref{suppeq:monomialexpansion}], it is sufficient to compute this integral for $\chi_{n,k}(\vec{r}) = \mathbf{p}(\vec{r})$, $\chi^+_{n^{\prime}, k^+}(x_1, \dots, x_{n^{\prime}}) = \mathbf{q}^+(x_1, \dots, x_{n^{\prime}})$, and $\chi^-_{n-n^{\prime}, k^-}(x_{n^{\prime}+1}, \dots, x_n) = \mathbf{q}^-(x_{n^{\prime}+1}, \dots, x_n)$, with the monomial exponents $\mathbf{p} = \{p_1,\dots,p_n\}$, $\mathbf{q}^+ = \{q^+_1,\dots,q^+_{n^{\prime}}\}$, and $\mathbf{q}^- = \{q^-_{n^{\prime}+1},\dots,q^-_n\}$, which yields
\begin{equation}
    \mathcal{C}^{\mathbf{p}}_{\mathbf{q}^+, \mathbf{q}^-} \pull= 
    \sqrt{n! \push n^{\prime}! \push (n-n^{\prime})!} \;
    \mathcal{I}_l \big( \{p_1 + q^+_1, \dots, p_{n^{\prime}} + q^+_{n^{\prime}} \}, \tilde{x} \big) \; 
    \mathcal{I}_r \big( \tilde{x}, \{p_{n^{\prime}+1} + q^-_{n^{\prime}+1}, \dots, p_n + q^-_n \} \big) \;,
\end{equation}
where $\mathcal{I}_l$ and $\mathcal{I}_r$ have closed-form expressions given in Eqs.~\eqref{suppeq:Il} and \eqref{suppeq:Ir}.

One can construct $\chi^{\pm}_{n,k}$ in terms of monomials using the same procedure as in Sec.~\ref{suppsec:construction}. For contact interactions, they can be generated from the eigenvalue equations $\big( \hat{K}^{\pm}_{\mathcal{L}} + c \push \hat{U}^{\pm}_{\mathcal{L}} \big) |\chi^{\pm}_{n,k}\rangle = \mathcal{E}^{\pm}_{n,k} |\chi^{\pm}_{n,k}\rangle$ where, in the position basis,
\begin{alignat}{2}
    \hat{K}^+_{\mathcal{L}} &= 
    - \frac{1}{2} \sum_{i} \frac{\partial}{\partial x_i} 
    \bigg[ x_i (\tilde{x} - x_i) \push \frac{\partial}{\partial x_i} \bigg] \;, \qquad 
    &&\hat{U}^+_{\mathcal{L}} = 
    \sum_{i < i^{\prime}} x_i (\tilde{x} - x_i) \push 
    \delta(x_i - x_{i^{\prime}}) \;,
    \\
    \hat{K}^-_{\mathcal{L}} &= 
    - \frac{1}{2} \sum_{i} \frac{\partial}{\partial x_i} 
    \bigg[ (x_i - \tilde{x}) (1 - x_i) \push \frac{\partial}{\partial x_i} \bigg] \;, \qquad 
    &&\hat{U}^-_{\mathcal{L}} = 
    \sum_{i < i^{\prime}} (x_i - \tilde{x}) (1 - x_i) \push 
    \delta(x_i - x_{i^{\prime}}) \;.
\end{alignat}
Note we have adapted the factors $x_i (1-x_i)$ in Eq.~\eqref{suppeq:basisgen} for the intervals $[0,\tilde{x}]$ and $[\tilde{x},1]$. Following the derivation in Eqs.~\eqref{suppeq:KL}--\eqref{suppeq:ULpq}, we find the matrix elements, for $\mathbf{p} = \{p_1,\dots,p_n\}$ and $\mathbf{q} = \{q_1,\dots,q_n\}$,
\begin{align}
    \big\langle \mathbf{p} \big| \hat{K}^+_{\mathcal{L}} \big| \mathbf{q} \big\rangle =&\; 
    \frac{n!}{2} \push \sum_{i=1}^n \push p_i q_i 
    \Big[ \tilde{x} \push \mathcal{I}_l \big( \texttt{incr}(\mathbf{p} + \mathbf{q}, i, -1), \tilde{x} \big) 
    - \mathcal{I}_l(\mathbf{p} + \mathbf{q}, \tilde{x}) \Big] \;,
    \\[0.1cm]
    \big\langle \mathbf{p} \big| \hat{U}^+_{\mathcal{L}} \big| \mathbf{q} \big\rangle =&\; 
    \frac{n!}{2} \push \sum_{i=1}^{n-1} \Big[\tilde{x} \push 
    \mathcal{I}_l \Big( \texttt{incr} \big( \texttt{merge}(\mathbf{p} + \mathbf{q},i), i, 1 \big), \tilde{x} \Big) - 
    \mathcal{I}_l \Big( \texttt{incr} \big( \texttt{merge}(\mathbf{p} + \mathbf{q},i), i, 2 \big), \tilde{x} \Big) \Big] \;,
    \\[0.1cm]
    \big\langle \mathbf{p} \big| \hat{K}^-_{\mathcal{L}} \big| \mathbf{q} \big\rangle =&\; 
    \frac{n!}{2} \push \sum_{i=1}^n \push p_i q_i 
    \Big[ (1+\tilde{x}) \push \mathcal{I}_r \big( \tilde{x}, 
    \texttt{incr}(\mathbf{p} + \mathbf{q}, i, -1) \big) 
    - \mathcal{I}_r(\tilde{x}, \mathbf{p} + \mathbf{q}) 
    - \tilde{x} \push \mathcal{I}_r \big( \tilde{x}, 
    \texttt{incr}(\mathbf{p} + \mathbf{q}, i, -2) \big) \Big] \;,
    \\[0.1cm]
    \nonumber
    \big\langle \mathbf{p} \big| \hat{U}^-_{\mathcal{L}} \big| \mathbf{q} \big\rangle =&\; 
    \frac{n!}{2} \push \sum_{i=1}^{n-1} \Big[(1+\tilde{x}) \push 
    \mathcal{I}_r \Big(\tilde{x}, \texttt{incr} \big( \texttt{merge}(\mathbf{p} + \mathbf{q},i), i, 1 \big) \Big) 
    - \mathcal{I}_r \Big(\tilde{x}, \texttt{incr} \big( \texttt{merge}(\mathbf{p} + \mathbf{q},i), i, 2 \big) \Big) 
    \\
    & \hspace{1.3cm} - \tilde{x} \push \mathcal{I}_r \big(\tilde{x}, \texttt{merge}(\mathbf{p} + \mathbf{q}, i) \big)\Big] \;.
\end{align}
The inner product of the symmetrized monomials over $[0,\tilde{x}]^n$ and $[\tilde{x},1]^n$ are simply given by $\langle \mathbf{p} | \mathbf{q} \rangle^+ \pull = n! \push \mathcal{I}_l (\mathbf{p} + \mathbf{q}, \tilde{x})$ and $\langle \mathbf{p} | \mathbf{q} \rangle^- \pull = n! \push \mathcal{I}_r (\tilde{x}, \mathbf{p} + \mathbf{q})$. Using these results, the construction of $\chi^{\pm}_{n,k}$ reduces to a generalized eigenvalue problem. In Fig.~\ref{fig:entanglement} of the main text, we truncate $\chi^{\pm}_{n,k}$ the same way as $\chi_{n,k}$ (as detailed in Sec.~\ref{suppsec:simul}).

\section{Tight-binding approximation with hard walls}
\label{suppsec:tightbinding}

\begin{figure}[!htb]
	\includegraphics[width=1\linewidth]{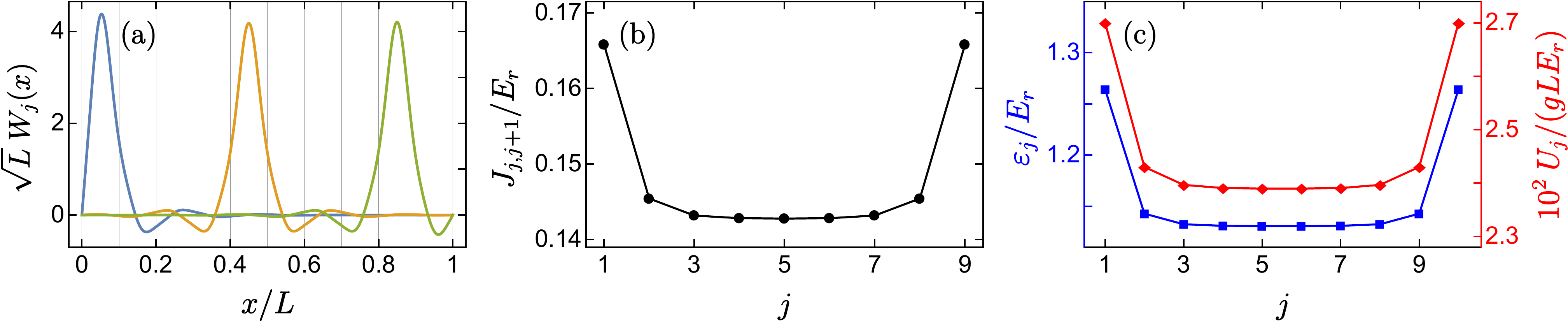}
	\centering
	\caption{(a) Wannier functions at the 1st, 5th, and 9th potential minima for $V_0/E_r=2$ and $N_w=10$. (b) Tunneling between nearest-neighbor minima and (c) local energy shifts $\varepsilon_i$ and interaction energies $U_i$ for the system in (a).
	}
	\label{suppfig:tightbinding}
\end{figure}

In the presence of a sufficiently deep external potential $V(x) = V_0 \cos^2 (N_w \pi x / L)$, one can approximate the continuum problem by $N_w$ localized Wannier orbitals at the potential minima. To derive this tight-binding model, we consider the single-particle Hamiltonian $\hat{H}_{\text{sp}} = \hat{K} + \hat{P}$, where $\hat{K} = -\partial_x^2/2$ and $\hat{P} = V(x)$ in the position basis. Since we have hard-wall boundaries at $x=0$ and $x=L$, the Hilbert space is spanned by the particle-in-a-box wave functions $\alpha_m (x) = \sqrt{2/L} \push \sin(m\pi x/L), \; m=1,2,3,\dots$, such that $\langle \alpha_{m} | \hat{K} | \alpha_{m^{\prime}} \rangle = \delta_{m,m^{\prime}} (m/N_w)^2 E_r$, where $E_r = N_w^2 \pi^2 / (2 L^2)$ is the recoil energy. The potential $V(x)$ couples these states with the amplitudes
\begin{equation}
    \langle \alpha_{m} | \hat{P} | \alpha_{m^{\prime}} \rangle = 
    (V_0/2) \push \delta_{m,m^{\prime}} + (V_0/4) \big(
    \delta_{m, m^{\prime} + 2N_w} + 
    \delta_{m, m^{\prime} - 2N_w} - 
    \delta_{m, -m^{\prime} + 2N_w} \big) \;,
\end{equation}
which vanish for $m \neq m^{\prime}$ unless $m$ and $m^{\prime}$ are separated by or add up to $2 N_w$. This selection rule partitions the wave functions into $N_w$ groups characterized by $q=1,2,\dots, N_w$, where $\text{mod}(m \pm q, 2N_w) = 0$. Here, $q$ plays the role of quasimomentum and the lowest-energy eigenstates of $\hat{H}_{\text{sp}}$ for each $q$ constitute the lowest band, $\hat{H}_{\text{sp}} |\psi_q \rangle = E_q |\psi_q \rangle$. We find the Wannier functions as eigenstates of $\hat{X}_{\text{proj}} = \hat{\Pi} \hat{X} \hat{\Pi}$, where $\hat{\Pi}$ is the projector onto the lowest band and $\hat{X}$ is the position operator with matrix elements
\begin{equation}
    \langle \alpha_m | \hat{X} | \alpha_{m^{\prime}} \rangle = 
    [(-1)^{m+m^{\prime}} \pull - 1] \push 
    \frac{4 L m m^{\prime}}{\pi^2 (m^2 - m^{\prime 2})^2} \;.
\end{equation}
Figure~\ref{suppfig:tightbinding}(a) shows the Wannier functions $W_j(x)$ for $N_w = 10$ and $V_0/E_r = 2$, centered at different potential minima $j$, which become more localized with increasing $V_0/E_r$. In this Wannier basis, one can calculate the nearest-neighbor tunneling $J_{j,j+1} = - \langle W_j | \hat{H}_{\text{sp}} | W_{j+1} \rangle$ and local energy shifts $\varepsilon_j = \langle W_j | \hat{H}_{\text{sp}} | W_j \rangle$. Contact interactions give rise to the ``on-site'' interaction energies $U_j = g \int_0^L \pull {\rm d}x \push |W_j(x)|^4$. As shown in Figs.~\ref{suppfig:tightbinding}(b-c), these effective Hubbard parameters are slightly larger close to the edges. In Fig.~\ref{fig:inhomogeneous} of the main text, we simulate such nonuniform Hubbard models using a standard DMRG routine in Mathematica version 12.3.0 with singular-value cutoff $\epsilon_{\text{trunc}} = 10^{-12}$, convergence threshold $\epsilon_{\text{conv}} = 10^{-8}$, and maximum bond dimension $D_{\text{max}} = 500$ (cf. Sec.~\ref{suppsec:simul}).

\section{Condensate fraction in the Tonks gas}
\label{suppsec:tonks}

In the limit of infinitely strong repulsive interactions, $\gamma \to \infty$, the 1D Tonks gas of impenetrable bosons maps onto free fermions, such that their ground-state wave function $\Psi(\vec{r})$ is given by the absolute value of that of the fermions, $\Psi(\vec{r}) = |\Psi_F(\vec{r})|$ \cite{supp_girardeau1960relationship}. For a sinusoidal potential $V(x) = V_0 \cos^2 (N_w \pi x / L)$ with unit filling, $N_w = N$, $\Psi_F(\vec{r})$ is obtained by populating each of the single-particle eigenstates $\psi_q (x)$ in the lowest band (see Sec.~\ref{suppsec:tightbinding}) with a fermion and taking the Slater determinant. Therefore,
\begin{equation}
    \Psi(\vec{r}) = \frac{1}{\sqrt{N!}} \push \Big| \text{det} 
    \big[ \psi_q(x_i) \big]_{q, i = 1,2,\dots,N} \Big| \;.
\end{equation}
The boson occupation of the single-particle modes are contained in one-body density operator $\hat{\rho}_1$, where
\begin{equation}
    \rho_1(x,x^{\prime}) = N \pull \int_0^L \pull\pull 
    {\rm d}^{N-1}r \; 
    \Psi(x, \vec{r}) \push \Psi(x^{\prime}, \vec{r}) \;.
\end{equation}
In particular, the condensate fraction is given by $f_0 \coloneqq N_0/N$, where $N_0$ is the occupation of the single-particle ground state (corresponding to $q=1$), $N_0 = \langle \psi_1 | \hat{\rho}_1 | \psi_1 \rangle$. Thus,
\begin{equation}
    f_0 = \frac{1}{N} \pull \int_0^L \pull\pull {\rm d}x \pull 
    \int_0^L \pull\pull {\rm d}x^{\prime} \; \psi_1(x) \push 
    \rho_1(x,x^{\prime}) \push \psi_1(x^{\prime}) \;.
\end{equation}
For $V_0=0$, $\psi_q(x) = \sqrt{2/L} \push \sin(q \pi x/L)$ and one can simplify $\rho_1(x,x^{\prime})$ to a determinant \cite{supp_forrester2003painleve}, reducing the calculation of $f_0$ to a 2D numerical integral. Figure~\ref{suppfig:vanishingpot}(a) shows that, in this case, $f_0 \sim N^{-0.45}$. The condensate fraction $f_0$ vanishes in the thermodynamic limit, but is finite for fixed $N$. For $V_0 > 0$, we find the single-particle states $|\psi_q\rangle$ by exact diagonalization and then compute $f_0$ by an $(N+1)$-dimensional Monte Carlo integration with up to $10^8$ points in Mathematica version 12.3.0. The results for $N=10$ are shown in Fig.~\ref{fig:inhomogeneous}(b) of the main text.

\section{Luttinger parameter and pinning instability}
\label{suppsec:pinning}

\begin{figure}[!htb]
	\includegraphics[width=1\linewidth]{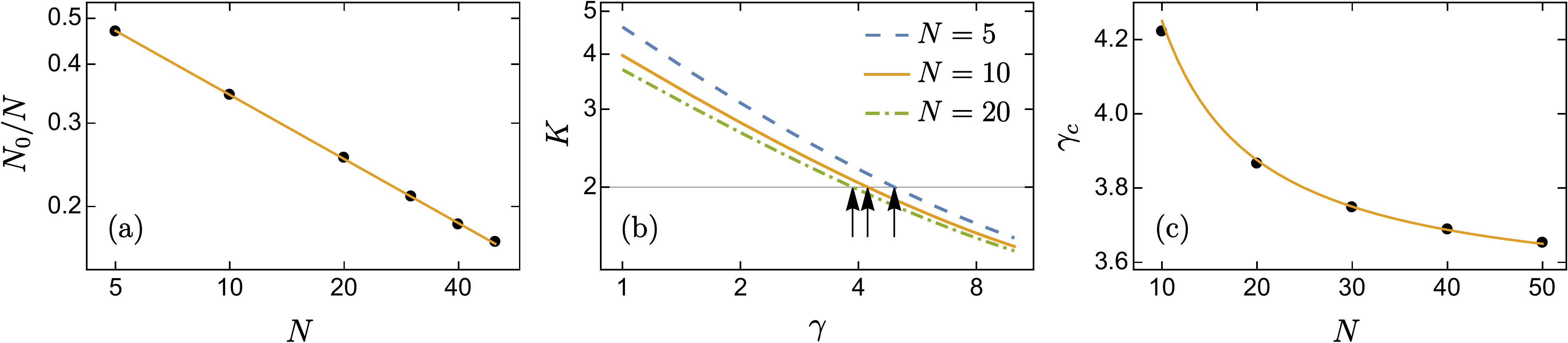}
	\centering
	\caption{Ground-state parameters for $V_0 \to 0$: (a) Condensate fraction vs particle number for $\gamma \to \infty$; solid line shows the fit $0.97 \push N^{-0.45}$. (b) Luttinger parameter vs interaction strength; arrows show the onset of pinning instability below $K=2$. (c) Critical interaction strength $\gamma_c$, corresponding to $K=2$, vs particle number; solid line shows the fit $3.5 + 7.5/N$.
	}
	\label{suppfig:vanishingpot}
\end{figure}

The low-energy excitations of our model for the interacting 1D Bose gas are described by a Luttinger liquid \cite{supp_cazalilla2004bosonizing}, characterized by the parameter $K=v_s/v_N$, where $v_s$ and $v_N$ are the speed of sound and density stiffness, respectively. These can be obtained from the ground-state energy $E$ as
\begin{equation}
    v_s = \sqrt{\frac{L^2}{N} \push \frac{\partial^2 E}{\partial L^2}} \quad \text{and} \quad
    v_N = \frac{L}{\pi} \push \frac{\partial^2 E}{\partial N^2} \;,
\end{equation}
with $\hbar=m=1$. In the absence of any external potential $V(x)$, $E$ can be calculated exactly using Bethe Ansatz \cite{supp_batchelor20051d}, thus one can find $K$ as a function of the interaction strength $\gamma$ for a given particle number $N$, as shown in Fig.~\ref{suppfig:vanishingpot}(b). Crucially, perturbative calculations have shown \cite{supp_buchler2003commensurate} that for $K<2$, a Luttinger liquid has an instability whereby it is pinned to an insulating state by an arbtrarily weak commensurate potential, $V(x) = V_0 \cos^2 (N_w \pi x/L)$ with $N/N_w =$ integer. Thus, $K=2$ marks the transition from a superfluid to a Mott insulator for $V_0 \to 0$. In Fig.~\ref{suppfig:vanishingpot}(c), we plot the corresponding interaction strength $\gamma_c$, which is well fitted by $\gamma_c \approx 3.5 + 7.5/N$.

\begingroup
\renewcommand{\addcontentsline}[3]{}
\renewcommand{\section}[2]{}


%

\endgroup

\end{document}